\documentclass[twocolumn]{aastex701}
\begin{document}

\title{A Wavelet-Integrated Search Pipeline for Narrowband Technosignatures in FAST Observations of 33 Exoplanet Systems}

\author{Zi-Qi Li}
\affiliation{Institute for Frontiers in Astronomy and Astrophysics, Beijing Normal University, Beijing 102206, China}
\affiliation{Institute for Astrophysics, School of Physics, Zhengzhou University, Zhengzhou 450001, China}
\affiliation{International College of Zhengzhou University, Zhengzhou 450001, China}
\email{ziqili@stu.zzu.edu.cn}  

\author{Jian-Kang Li}
\affiliation{Institute for Frontiers in Astronomy and Astrophysics, Beijing Normal University, Beijing 102206, China}
\affiliation{Department of Astronomy, Beijing Normal University, Beijing 100875, China}
\email{lijiankang@mail.bnu.edu.cn}  

\author{Zhe-Wei Luo}
\affiliation{School of Computer and Artificial Intelligence, Zhengzhou University, Zhengzhou 450001, China}
\email{zwluo@stu.zzu.edu.cn}

\author{Chen-Xu Guan}
\affiliation{International College Of Zhengzhou University, Zhengzhou 450001, China}
\email{cxguan@stu.zzu.edu.cn}

\author{Yu-Tong Fan}
\affiliation{International College Of Zhengzhou University, Zhengzhou 450001, China}
\email{ytfan@stu.zzu.edu.cn}

\author{Zhen-Zhao Tao}
\affiliation{Institute for Frontiers in Astronomy and Astrophysics, Beijing Normal University, Beijing 102206, China}
\affiliation{Department of Astronomy, Beijing Normal University, Beijing 100875, China}
\email{taozhenzhao@dzu.edu.cn}

\author{Xiao-Hang Luan}
\affiliation{Institute for Frontiers in Astronomy and Astrophysics, Beijing Normal University, Beijing 102206, China}
\affiliation{Department of Astronomy, Beijing Normal University, Beijing 100875, China}
\email{xhluan@mail.bnu.edu.cn}

\author{Bo-Lun Huang}
\affiliation{Institute for Frontiers in Astronomy and Astrophysics, Beijing Normal University, Beijing 102206, China}
\affiliation{Department of Astronomy, Beijing Normal University, Beijing 100875, China}
\affiliation{School of Physics and Astronomy, University of Glasgow, Glasgow G12 8QQ, United Kingdom}
\email{bolunh@hotmail.com}

\author{Yu Hu}
\affiliation{Institute for Frontiers in Astronomy and Astrophysics, Beijing Normal University, Beijing 102206, China}
\affiliation{Department of Astronomy, Beijing Normal University, Beijing 100875, China}
\affiliation{National Astronomical Observatories, Chinese Academy of Sciences, Beijing 100012, China}
\email{202431101094@mail.bnu.edu.cn}

\author{Peng-Yu Li}
\affiliation{International College Of Zhengzhou University, Zhengzhou 450001, China}
\email{lipengyu87020309@stu.zzu.edu.cn}

\author{Pu-Fan Liu}
\affiliation{Institute for Astrophysics, School of Physics, Zhengzhou University, Zhengzhou 450001, China}
\affiliation{International College Of Zhengzhou University, Zhengzhou 450001, China}
\email{liupufan2022@163.com}

\author{Kang Jiao}
\affiliation{Institute for Astrophysics, School of Physics, Zhengzhou University, Zhengzhou 450001, China}
\email{kangjiao@zzu.edu.cn}

\author{Tong-Jie Zhang\thanks{Corresponding author}}
\affiliation{Institute for Frontiers in Astronomy and Astrophysics, Beijing Normal University, Beijing 102206, China}
\affiliation{Department of Astronomy, Beijing Normal University, Beijing 100875, China}
\affiliation{Institute for Astronomical Science, Dezhou University, Dezhou 253023, China}
\email[show]{tjzhang@bnu.edu.cn}  

\author{Hai-Yan Zhang}
\affiliation{National Astronomical Observatories, Chinese Academy of Sciences, Beijing 100012, China}
\email{hyzhang@nao.cas.cn}

\author{Peng Jiang}
\affiliation{National Astronomical Observatories, Chinese Academy of Sciences, Beijing 100012, China}
\email{pjiang@bao.ac.cn}

\author{Rui Li}
\affiliation{Institute for Astrophysics, School of Physics, Zhengzhou University, Zhengzhou 450001, China}
\email{liruiww@zzu.edu.cn}

\author{Liang Gao}
\affiliation{Institute for Frontiers in Astronomy and Astrophysics, Beijing Normal University, Beijing 102206, China}
\affiliation{Institute for Astrophysics, School of Physics, Zhengzhou University, Zhengzhou 450001, China}
\email{liruiww@zzu.edu.cn}
1
% \author[orcid=0000-0000-0000-0001,sname='xxx']{xxx}
% \altaffiliation{xxx}
% \affiliation{xxx}
% \email[show]{xxx}  

%\collaboration{all}{}

%%%%%%%%%%%%%%%%%%%%%%%%%%%%%%%%%%%%%%%%%%%%%%%%%%%%%%%%%%%%%%%%%%%%%%%%%%%%%%%%
\begin{abstract}
Building on prior FAST targeted and blind SETI campaigns toward 33 exoplanet systems, we introduce a wavelet-integrated search pipeline for narrowband technosignature candidates in radio dynamic spectra. At its core, the pipeline uses a Multi-Scale Wavelet Net (MSWNet) to produce an interpretable multi-resolution representation, followed by a lightweight parameter estimator for endpoint localization. Rather than relying solely on hard-threshold drift searches, the pipeline reframes narrowband detection as wavelet-guided feature extraction followed by endpoint regression, morphology-aware filtering, raw-data S/N validation, and multi-beam anticoincidence veto. Applied to real FAST data, the pipeline recovers representative events from prior analyses and produces a compact set of veto-ready candidates for downstream inspection. The resulting workflow preserves interpretability, low regression complexity, and auditable threshold control, making it readily transferable to other radio surveys and large-scale technosignature searches.
\end{abstract}

%\uat{Technosignatures}{2128}

\section{Introduction}
\label{sec:intro}
The search for extraterrestrial intelligence (SETI)\citep{1959Natur.184..844C,1961PhT....14d..40D} targets one of humanity’s oldest scientific questions—whether intelligent life exists beyond Earth—by prioritizing technosignatures, whose engineered nature can make them more directly testable than faint biosignatures \citep{2014PNAS..11112634S}. In radio SETI, the most widely pursued targets are narrowband drifting signals, since natural astrophysical processes rarely generate extremely narrow spectral features while engineered transmitters often do\citep{2001ARA&A..39..511T}. Over decades, ground-based searches have progressed from narrow fixed-frequency scans to wideband, high-resolution surveys enabled by modern digital backends\citep{2018PASP..130d4502M}, translating sensitivity and bandwidth gains into surging candidate volumes\citep{anderson2002seti,2020AJ....159...86P}. As a result, verification—rather than raw detection—has become the dominant bottleneck\citep{2020AJ....159...86P}, particularly under pervasive terrestrial RFI that can mimic narrow drifting tracks\citep{2010MNRAS.405..155O,2022MNRAS.516.5367M,2020AJ....159...86P}.

Within this landscape, the Five-hundred-meter Aperture Spherical radio Telescope (FAST)\citep{2011IJMPD..20..989N,2018IMMag..19..112L,2020Innov...100053Q} provides exceptional sensitivity, sky coverage, and 19-beam capability in the L band (1.05–1.45 GHz)\citep{2019SCPMA..6259502J}, enabling both commensal and dedicated SETI observations\citep{2020RAA....20...78L,2095-7777(2020)02-0158-06}. Critically, the multibeam receiver supports simultaneous spatial discrimination—using non-primary beams as contemporaneous OFF-source references—forming the basis of robust multi-beam coincidence matching (MBCM)\citep{2005RaSc...40.5S18H}. FAST’s early infrastructure demonstrated commensal data collection with fine spectral resolution and strong RFI rejection efficiency \citep{2095-7777(2020)02-0158-06}, and subsequent work established FAST+MBCM as a scalable validation paradigm in both targeted and blind search settings\citep{2022AJ....164..160T,2023AJ....165..132L,2023AJ....166..190T,2020AJ....159...86P}.

At the algorithmic level, most narrowband pipelines still operate on STFT-derived dynamic spectra\citep{2001SPIE.4273..104W} and Doppler drift-rate searches, from early large-scale efforts (e.g., SERENDIP/SETI@home) to modern coherent tree-based implementations such as \texttt{TurboSETI}\citep{2001SPIE.4273..104W,anderson2002seti,2013ApJ...767...94S,2017ApJ...849..104E,2020AJ....159...86P}. These baselines remain effective, but two practical constraints dominate in real deployments. First, under complex anthropogenic RFI, drift searches can produce large numbers of spurious detections (i.e., our pipeline yields up to $\sim 2.7\times10^6$ detections under conservative settings in Figure~\ref{fig:filtering}) that translate directly into expensive human review cycles\citep{2010MNRAS.405..155O,2022MNRAS.516.5367M,2020AJ....159...86P}. Second, the commonly assumed strictly linear drift model can reduce sensitivity to more complex Doppler profiles induced by realistic transmitter/receiver dynamics\citep{2016AJ....152..181H,2020AJ....159...86P}. In short, the “detection → inspection” loop scales poorly when RFI is non-stationary and visually diverse in the time–frequency plane.

A complementary design principle is therefore to introduce an inductive bias that matches the structure of time–frequency data. Wavelets provide an explicit multi-scale decomposition that separates coarse structures from fine details, yielding an interpretable representation in which narrowband tracks and broadband/block-like interference can be disentangled across subbands\citep{9785993}. This is particularly attractive for SETI dynamic spectra—time–frequency diagrams derived from Filterbank files—where weak narrow drifting lines are entangled with morphologically variable, non-stationary RFI in the same plane\citep{2010MNRAS.405..155O, 8918798, 2022MNRAS.516.5367M}. Recent radio astronomy practice increasingly reframes RFI mitigation and signal recovery as learnable image-like problems on dynamic spectra (e.g., segmentation/denoising)\citep{2017A&C....18...35A, 2022MNRAS.516.5367M, 2021RAA....21..119Y, ZHANG2024100822, 2025A&A...701A..81Y}. Meanwhile, mature detection ideas from computer vision (including detector-style post-processing such as NMS) have begun to migrate into astronomy, suggesting that when target morphology is salient, parameter regression and localization can leverage established detection practice\citep{2023A&A...677A.101G,10.3389/fspas.2025.1656917,2024A&A...690A.211C,1699659,8237855,7780460}. However, when such designs act directly on raw observations, performance can be constrained by measurement noise and non-stationary interference, limiting robustness\citep{2017MNRAS.467.1661V,ZHANG2024100822}. This motivates hybrid schemes that couple a structured, auditable front-end (e.g., wavelet-consistent representations) with lightweight learning modules for complex morphology discrimination\citep{2025A&A...701A..81Y, ZHANG2024100822, 9785993, 2021RAA....21..119Y, 8918798}.

Here we present an inference-time pipeline for targeted SETI with FAST, converting high-volume time-frequency data into a compact, veto-ready candidate stream via wavelet-consistent cleaning (MSWNet) and lightweight endpoint regression, followed by auditable staged filtering and 19-beam anticoincidence veto. Real-data results from the 33-target campaign are in Section~\ref{sec:results}; notable events, trade-offs, and comparisons to prior work are in Section~\ref{sec:discussion}; full methods are in Section~\ref{sec:methods}.

\section{Proposed Methods}
\label{sec:methods}

\subsection{Data Simulation}
\label{sec:data-sim}
Our search builds on FAST L-band (1.05--1.45~GHz) observations of 33 exoplanet systems conducted in 2021 using the 19-beam receiver in spectral-line mode, providing dynamic spectra at $\Delta\nu\simeq7.5$~Hz and $\Delta t=10$~s \citep{2022AJ....164..160T,2023AJ....165..132L}. Candidate mining and validation follow the established multi-beam anticoincidence (MBCM) framework, treating the tracked central beam as ON-source and the remaining beams as simultaneous OFF-source veto channels \citep{2005RaSc...40.5S18H,2022AJ....164..160T,2023AJ....165..132L}.
To provide high-quality training data and reproduce the patterns of FAST data faithfully, we simulate dynamic spectra in the FAST L-band detection mode using a clean 1278–1380~MHz background segment identified by Luan et al. (2023) as relatively free of persistent RFI~\citep{2023AJ....165..132L}. Starting from this real noise bed, we inject synthetic technosignature signals using a custom pipeline based on the open-source setigen library~\citep{2022AJ....163..222B}. Two signal morphologies are introduced: (1) constant-drift tracks, which appear as near-linear diagonal lines in the time–frequency plane, and (2) accelerating tracks with quadratic frequency drift (i.e., parabolic curves). The yielded signals are probabilistically modulated in time—via pulsed or sinusoidal envelopes—and in frequency, increasing the diversity of the injected morphologies. We draw the intrinsic spectral width of injected technosignatures uniformly from 1--3 frequency channels (i.e., $W\in[1,3]\Delta\nu$), representing unresolved to mildly broadened narrowband emitters under our channel resolution. All simulated signals respect an absolute drift-rate limit of $|\dot{\nu}|=4.0~\text{Hz/s}$~\citep{2026AJ....171...51H}, consistent with FAST’s coherent search range. We derive the minimum resolvable drift as
\begin{equation}
\dot{\nu}_{\min} = \frac{\Delta\nu}{N_t \Delta t}
\end{equation}
\citep{2024AJ....167....8L}, where $\Delta\nu$ is the frequency channel width and $N_t \Delta t$ is the total integration time. This $\dot{\nu}_{\min}$ sets a lower bound below which a signal’s frequency change over the observation is less than one channel and effectively indistinguishable from $\dot{\nu}=0$. To balance input size and drift coverage, we fix the frequency resolution to $N_f=256$. Given FAST's L-band channel width $\Delta\nu \simeq 7.5\,\mathrm{Hz}$ and the survey drift-rate bound
${\rm MDR}=4\,\mathrm{Hz/s}$ \citep{2022AJ....164..160T,2023AJ....165..132L}, one patch spans
$\Delta\nu_{\rm patch}=N_f\Delta\nu \approx 1.9\,\mathrm{kHz}$, i.e.,
$T_{\max}\approx \Delta\nu_{\rm patch}/{\rm MDR}\approx 480\,\mathrm{s}$ for the worst-case drift. A 2.5\% guard band is reserved on the side toward which the signal drifts, preventing the trajectory from crossing the patch boundary.

We use setigen to synthesize RFI-like contaminants by injecting nearly non-drifting narrowband signals whose widths are broadened to $\sim$1--10 times the channel width, with a probability of temporal modulation (like the synthetic technosignature signals). This encourages the model to distinguish genuinely drifting tracks from quasi-stationary RFI. In addition, we lightly enrich the background with hand-crafted stochastic features—short-lived narrowband spots and broader transient fluctuation bands—mimicking weak, time-variable instrumental or RFI residuals and improving noise robustness. Each simulated training box comprises a noisy spectrum (background + noise bands + RFI + synthetic signals), a corresponding clean spectrum (signals only), and a set of ground truth (GT) annotations $GT Boxes$. Each ground truth entry encodes a signal instance in the format $(\mathrm{class}, f_{\mathrm{start}}, f_{\mathrm{stop}})$, where $f_{\mathrm{start}}$ and $f_{\mathrm{stop}}$ are the frequencies at the beginning and end of the observation (i.e., at first and last time channel), and $\mathrm{class}$ indicates the signal type. We assign $\mathrm{class}=1$ to near-linear (constant-drift) signals and $\mathrm{class}=0$ to other signals. All ground-truth objects are assigned a target confidence (objectness label) of 1.0, following standard object-detection training practice \citep{7485869}.

%frequency acceleration --> drift acceleration
In practice there is a continuum of curvature, so we treat weakly accelerating tracks as linear.
Signals injected with setigen can be written in a general polynomial form in channel--time coordinates,
\begin{equation}
    c_n = c_0 + \sum_{p=1}^{P} a_p n^p, \qquad n=0,\dots,T-1,
\end{equation}
where higher-order terms capture increasing curvature complexity. In practice, setigen implements a quadratic (parabolic) approximation in our simulations, i.e., $c_n = c_0 + s\,n + a_2 n^2$. In this quadratic truncation, $a_2$ controls curvature in channel--time coordinates. The corresponding drift acceleration is
\begin{equation}
    \ddot{\nu} = \frac{2 a_2 \,\Delta \nu}{\Delta t^2}.
\end{equation}
We label a track as ``linear'' (class~1) if its maximum deviation from the best-fit straight line over the full observation is below $0.2$ channels. This is satisfied for $|a_2|\le 7.5\times10^{-5}$, corresponding to
$|\ddot{\nu}| \simeq 1.1\times10^{-5}\,\mathrm{Hz/s^2}$ and $\Delta \nu_{\max}\simeq 1.2\,\mathrm{Hz}\approx 0.16$ channels.

Physically, a narrowband transmitter on a rotating, orbiting planet should exhibit a smooth Doppler drift that is well approximated as linear over typical SETI snapshots (tens of seconds); rotation and orbital motion introduce only slowly varying, quasi-sinusoidal drift-rate changes on hour--day timescales, making curvature negligible within a single integration \citep{2022ApJ...938....1L}. In contrast, appreciable curvature over such short spans implies unusually large line-of-sight accelerations (e.g., an accelerating spacecraft or an actively frequency-adjusted transmitter), scenarios generally treated as rare when defining physically motivated drift-rate bounds \citep{2019ApJ...884...14S}. 

\subsection{Model Structure}
\label{sec:architecture}

\begin{figure*}[htbp]
\centering
\includegraphics[width=\linewidth]{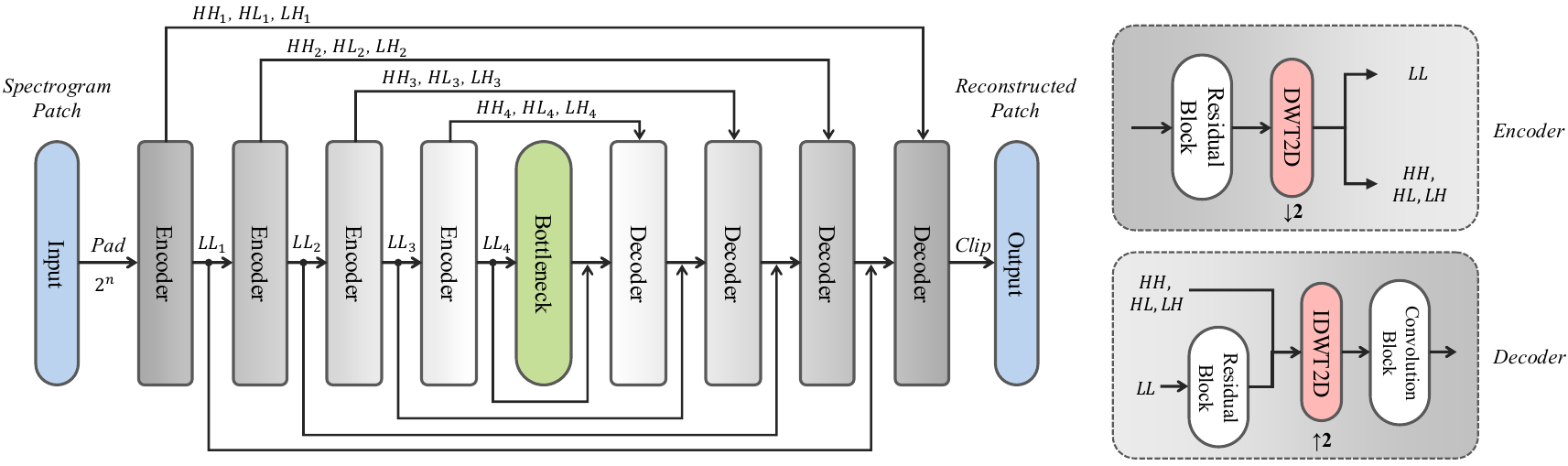}
\caption{Schematic architecture of MSWNet. The overall encoder–decoder structure is summarized in the left panel, while the right panel expands a single stage showing the internal encoder/decoder layout, where all resolution changes are performed within the blocks. The encoder replaces pooling with \texttt{DWT2D}, caching detail coefficients (LH, HL, HH) at each level, which are then reused by the decoder through \texttt{IDWT2D} together with refined coarse (LL) representations. Features in the bottleneck are processed by three depthwise-separable convolution blocks without change in resolution or dimensionality.}
\label{fig:mswnet}
\end{figure*}

Our main model, dubbed MSWNet (Multi-Scale Wavelet Net), is designed as an interpretable feature-extraction and reconstruction module (see Figure~\ref{fig:mswnet}). It consists of a wavelet-based encoder--decoder that performs multi-scale spectral reconstruction and yields physically consistent representations, while a lightweight module handles downstream candidate generation through parameter estimation and morphology-based filtering. The core encoder--decoder follows a U-Net design, but replaces conventional pooling with a two-dimensional discrete wavelet transform (\texttt{DWT2D}) at each scale. By reusing the encoder-derived wavelet detail coefficients, the decoder constrains fine-scale reconstruction and suppresses spurious high-frequency artifacts, so the network mainly learns to adjust coarse spectral structure while preserving localized narrowband tracks. Specifically, each encoder stage applies \texttt{DWT2D} to the current feature map, producing one approximation band (LL) and three detail bands (LH/HL/HH). The LL band is forwarded to the next stage, while the corresponding detail coefficients are cached and later reused by the decoder through inverse 2D discrete wavelet transform (\texttt{IDWT2D}), so that fine-scale content is carried through the reconstruction path rather than being hallucinated. In addition, LL features are propagated via skip concatenations to preserve contextual information across scales. We use residual blocks at each scale, and employ depthwise-separable convolution blocks in the bottleneck to reduce parameter and computation costs while preserving local pattern modeling capacity.

On the decoder side, we progressively reconstruct the spectrogram via inverse wavelet transforms. At each stage, the upsampled approximation features (LL) are refined and concatenated with the corresponding encoder-side LL skip features, and \texttt{IDWT2D} then merges them with the cached encoder-domain detail coefficients (LH/HL/HH) to recover the next higher resolution. Formally,
\begin{equation}
S^{(k-1)} = W^{-1}\!\left(LL_{\mathrm{dec}}^{(k)},\, H_{\mathrm{enc}}^{(k)}\right),
\end{equation}
where $H_{\mathrm{enc}}^{(k)}$ collects the encoder-side detail bands at scale $k$ and $LL_{\mathrm{dec}}^{(k)}$ is the decoder-refined approximation band.

Downstream of MSWNet, we employ a lightweight parameter estimator to convert the reconstructed feature map into candidate predictions. This estimator is intentionally decoupled from MSWNet: MSWNet acts as a physically constrained feature extractor, while the downstream module performs only low-cost parameter estimation and subsequent morphology-based filtering, improving interpretability, efficiency, and robustness across observing conditions. Concretely, the estimator preserves the full frequency axis throughout the convolutional backbone (i.e., no downsampling along $N_f$), ensuring that $f_{\mathrm{start}}$ and $f_{\mathrm{stop}}$ can be localized with maximal frequency precision. In contrast, the time axis is compressed to two bins via adaptive pooling, yielding a $2\times N_f$ (or $2\times F'$) representation. Using 2 (rather than 1) time bins is essential: the two rows retain start-versus-end information so that drift direction and magnitude remain identifiable, matching the $(f_{\mathrm{start}}, f_{\mathrm{stop}})$ annotation format. The estimator regresses a fixed set of $N$ predictions, outputting confidence, $(f_{\mathrm{start}}, f_{\mathrm{stop}})$, and class logits.

\subsection{Pipeline}
\label{sec:pipeline}
\begin{figure*}[htbp]
\centering
\includegraphics[width=\linewidth]{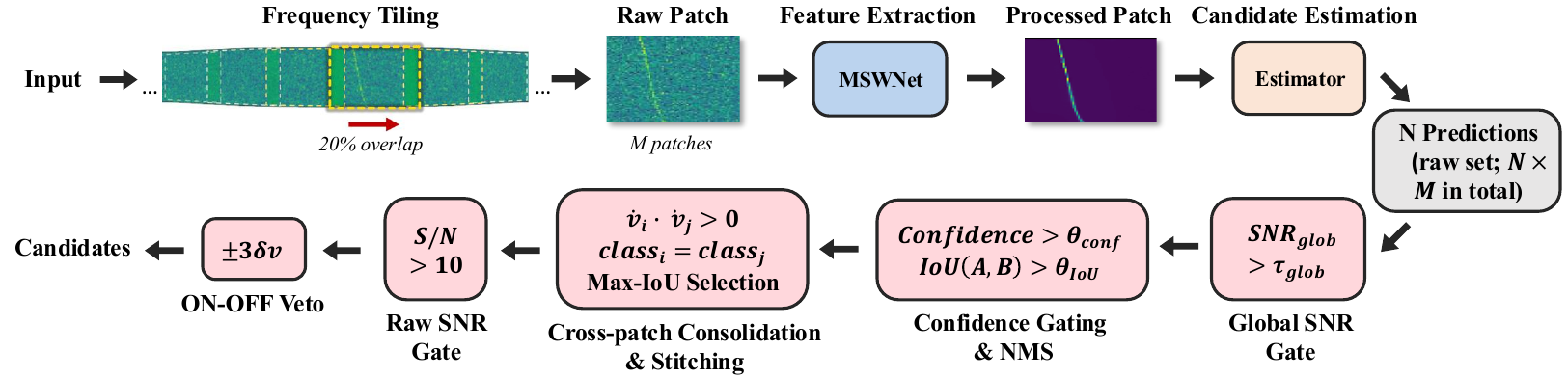}
\caption{Inference pipeline. MSWNet transforms each raw patch into a processed patch (cleaned map), and a lightweight parameter estimator outputs a fixed set of predictions. Predictions are gated by a patch-scale global SNR statistic, filtered by confidence and IoU (NMS), stitched across overlapping patches, and then validated by S/N measured on the raw spectrogram. Events surviving the veto stage constitute the final candidate set.}
\label{fig:pipeline}
\end{figure*}

At inference time, we deploy the method as a staged signal-search pipeline (Figure~\ref{fig:pipeline}) that converts a raw spectrogram into a final list of high-confidence candidate events.

% pipeline overview figure

\subsubsection{Pre-training objective}
While the pipeline in radio SETI mainly refers to inference-time search, the machine learning components must be pretrained beforehand to output reliable candidate sets for that stage. We train the lightweight parameter estimator in a set-based manner. Given an input spectrogram, MSWNet produces a reconstructed (cleaned) representation, and the estimator outputs a fixed set of $N$ predictions per sample, $\{\hat{b}_i\}_{i=1}^{N}$, where each $\hat{b}_i$ contains $(f_{\mathrm{start}}, f_{\mathrm{stop}}, \mathrm{class}, \mathrm{confidence})$. Since the output is an unordered set, we align predictions to ground truth via optimal bipartite matching and train with a matching-based loss.

Let $\{b_j\}_{j=1}^{M}$ be the ground-truth boxes. We compute the minimum-cost assignment using the Hungarian algorithm with the matching cost
\begin{equation}
C_{ij} = 1 - \mathrm{IoU}(\hat{b}_i,b_j) + 0.5\,\Delta_{\mathrm{class}}(\hat{b}_i,b_j) + 0.5\,\Delta_{\mathrm{dir}}(\hat{b}_i,b_j).
\end{equation}
Here $\mathrm{IoU}(\hat{b}, b)$ is the intersection-over-union between two frequency--time boxes (each signal is represented by its start/stop frequency span over the full time extent). $\Delta_{\mathrm{class}}$ is 0 if the predicted and true classes agree and 1 otherwise. $\Delta_{\mathrm{dir}}$ penalizes drift-direction reversal: it is 0 if the ordering of $f_{\mathrm{start}}$ and $f_{\mathrm{stop}}$ matches the truth and 1 if the sign is inverted. This term is important because upward and downward drifts are not physically interchangeable; an incorrect sign changes the Doppler interpretation and can invalidate subsequent analysis. In SETI analyses, drift direction is informative about the sign of the line-of-sight relative acceleration between a transmitter and the observer \citep{2017ApJ...849..104E}. For example, the Breakthrough Listen candidate BLC1 exhibited a positive drift rate of $+0.038~{\rm Hz/s}$ \citep{2021NatAs...5.1148S}; follow-up verification showed that its drift behavior is not consistent with the expected sidereal/barycentric drift along the Proxima Centauri pointing (with Earth’s rotation as a major contributor), and discussed the possibility of transmitter-side frequency compensation \citep{2021NatAs...5.1153S}. Incorporating $\Delta_{\mathrm{dir}}$ therefore prevents inverted-drift predictions from being matched to opposite-direction ground truth even when their frequency spans overlap.

We then minimize a composite loss enforcing localization, regression, classification, and confidence:
\begin{eqnarray}
\mathcal{L}
&=&
\sum_{(i,j)\in\mathcal{M}}
\Bigl[
\lambda_{\mathrm{giou}}
\left(1-\mathrm{gIoU}(\hat{b}_i,b_j)\right)
+\lambda_{\mathrm{reg}}
\left\|\hat{\mathbf{r}}_i-\mathbf{r}_j\right\|_2^2
\nonumber
+\lambda_{\mathrm{cls}}\,\mathrm{CE}(\hat{y}_i,y_j)
\Bigr]
+\lambda_{\mathrm{neg}}
\sum_{i\notin\mathcal{M}}
\mathcal{L}_{\mathrm{neg}}(\hat{p}_i).
\end{eqnarray}
where $\mathcal{M}$ is the set of matched prediction--ground-truth pairs returned by the Hungarian assignment; $\mathrm{gIoU}$ denotes generalized IoU (IoU with a non-overlap penalty); $\mathrm{CE}$ denotes cross-entropy on the class label; and $\mathbf{r}=(f_{\mathrm{start}},f_{\mathrm{stop}},p)$ with $p=1$ for matched targets. Unmatched predictions are treated as negatives, using a confidence regularizer $\mathcal{L}_{\mathrm{neg}}$ (e.g., focal loss) on $\hat{p}_i$.

\subsubsection{Inference}
Given a test spectrogram, we process it in frequency-overlapping patches and then apply staged post-processing to obtain the final hit list. MSWNet produces a cleaned map for each patch, and the parameter estimator outputs an initial set of candidate predictions $\{\hat{b}_i\}$ with confidence, $(f_{\mathrm{start}}, f_{\mathrm{stop}})$, and class logits.

FAST targeted observations are short ($\sim$20~min for most targets; $\sim$4~min for HD-111998) with $\Delta t=10$~s and $\Delta\nu\simeq 7.5$~Hz; therefore we do not tile along time and only tile along frequency with overlap ratio $r=0.2$. With $\Delta\nu_{\mathrm{patch}}=N_f\Delta\nu$ and $\Delta\nu_{\mathrm{ov}}=r\,\Delta\nu_{\mathrm{patch}}$, a sufficient condition for a drifting narrowband track to remain observable across adjacent frequency patches is
\begin{equation}
{\rm MDR}\,\Delta t + W + 2\epsilon_\nu \le \Delta\nu_{\mathrm{ov}},
\end{equation}
where $W$ is the effective track width (upper-bounded by our injection setting) and $\epsilon_\nu$ is a conservative frequency-localization margin. We take $\epsilon_\nu=3\Delta\nu$, which is the tolerance used in multibeam RFI matching \citep{2022AJ....164..160T}. For $N_f=256$, $\Delta\nu_{\mathrm{patch}}\approx 1.92$~kHz and $\Delta\nu_{\mathrm{ov}}\approx 384$~Hz, giving ${\rm MDR}_{\max}\approx(\Delta\nu_{\mathrm{ov}}-(W+2\epsilon_\nu))/\Delta t \approx 32~\mathrm{Hz/s}$ (with $W\simeq 3\Delta\nu$), well above the survey bound ${\rm MDR}=4~\mathrm{Hz/s}$ \citep{2022AJ....164..160T,2023AJ....165..132L}. For typical targets ($T_{\rm obs}\approx 20~\mathrm{min}$), this overlap budget implies a no-stitch completeness bound of ${\rm MDR}\lesssim(\Delta\nu_{\mathrm{ov}}-(W+2\epsilon_\nu))/T_{\rm obs}\approx 0.26~\mathrm{Hz/s}$, within which tracks are guaranteed to be fully contained in at least one patch and can be recovered without any stitching. For signals beyond this regime, stitching is applied conservatively, and our final results also indicate that most events are recovered under the no-stitch regime while stitch-required cases are rare.

Because this raw set may contain duplicates (multiple predictions on the same bright signal) and occasional noise triggers, we apply the following post-processing:

\begin{enumerate}

% global SNR filtering (robust activation gate).
\item To eliminate obvious noise triggers before fine validation, we compute a robust ``global SNR’’ score for each patch. Let $x\in\mathbb{R}^{n}$ be the vector of pixel intensities in the cleaned map. We estimate a robust location and scale using the median and MAD: $m=\mathrm{median}(x)$ and $\hat{\sigma}=1.4826\cdot \mathrm{median}\bigl(|x-m|\bigr)$. We then compute a robust peak statistic as the mean of the top-$k$ intensities, with $k=\max\!\bigl(k_{\min},\,\lfloor \rho n \rfloor\bigr)$ and $\mu_{\mathrm{top}}=\frac{1}{k}\sum_{u\in \mathrm{TopK}(x,k)} u$, and define
\begin{equation}
\mathrm{SNR}_{\mathrm{glob}}=\frac{\mu_{\mathrm{top}}-m}{\hat{\sigma}}.
\end{equation}
Patches with $\mathrm{SNR}_{\mathrm{glob}}<\tau_{\mathrm{glob}}$ are rejected (see Section~\ref{sec:results}). Compared to a single-pixel maximum test, this MAD-normalized top-$k$ statistic is less sensitive to isolated outliers and provides a stable activation gate for suppressing noise-dominated responses. In all experiments, $(\rho, k_{\min}, \tau_{\mathrm{glob}})$ are selected once on a held-out validation set and then fixed for all test observations; because $\mathrm{SNR}_{\mathrm{glob}}$ is MAD-normalized, the same threshold transfers reliably across varying noise floors.

% Confidence gating and overlap suppression
\item For each surviving patch, we discard predictions with confidence below a cutoff, and then apply non-maximum suppression (NMS): any lower-confidence prediction overlapping a higher-confidence one beyond an IoU threshold is removed, leaving at most one patch-level detection per signal (discussed in Section~\ref{sec:results}).

% Cross-patch consolidation and stitching
\item After patch-level inference, coordinates of detections are mapped back and duplicates across neighboring patches are merged by IoU-based suppression. Boundary-truncated segments are stitched only when drift direction/morphology are consistent and the overlap region is sparsely populated; otherwise stitching is skipped to avoid erroneous merges.

% S/N validation on raw data
\item Next, we compute a per-detection SNR using the raw (noisy) spectrogram. For each detection, we extract pixels along the inferred track defined by $(f_{\mathrm{start}}, f_{\mathrm{stop}})$ (and the assumed drift shape) and estimate the background from nearby off-track frequencies, following standard S/N measurements in classical pipelines \citep{2026AJ....171...51H}. We apply a SNR cutoff (e.g., ${\rm S/N}>10$), yielding a list of high-confidence events characterized by $(f_{\mathrm{start}}, f_{\mathrm{stop}})$, class, confidence, and measured S/N.

% Multi-beam anticoincidence (on-off veto)
\item Finally, for each surviving event, we check all non-primary beams for coincident events within the prescribed tolerance; events exhibiting multi-beam coincidence are rejected, and only Beam-1-localized events are retained as candidates.
\end{enumerate}

A pipeline code implementation is provided in \texttt{WaveSETI}\footnote{Code available at \url{https://github.com/Riko-Neko/waveseti}.}, implemented in \texttt{PyTorch} with runtime-profile configuration, data generation using \texttt{blimpy} and \texttt{setigen}, \texttt{.fil}/\texttt{.h5} loaders, Filterbank inference, an optional \texttt{PyQt5}-based pipeline visualization user interface (UI), candidate filtering, and post-processing utilities (including veto).

\section{Results}
\label{sec:results}

\subsection{Evaluation and results}

\begin{figure*}[htbp]
\centering
\includegraphics[width=\linewidth]{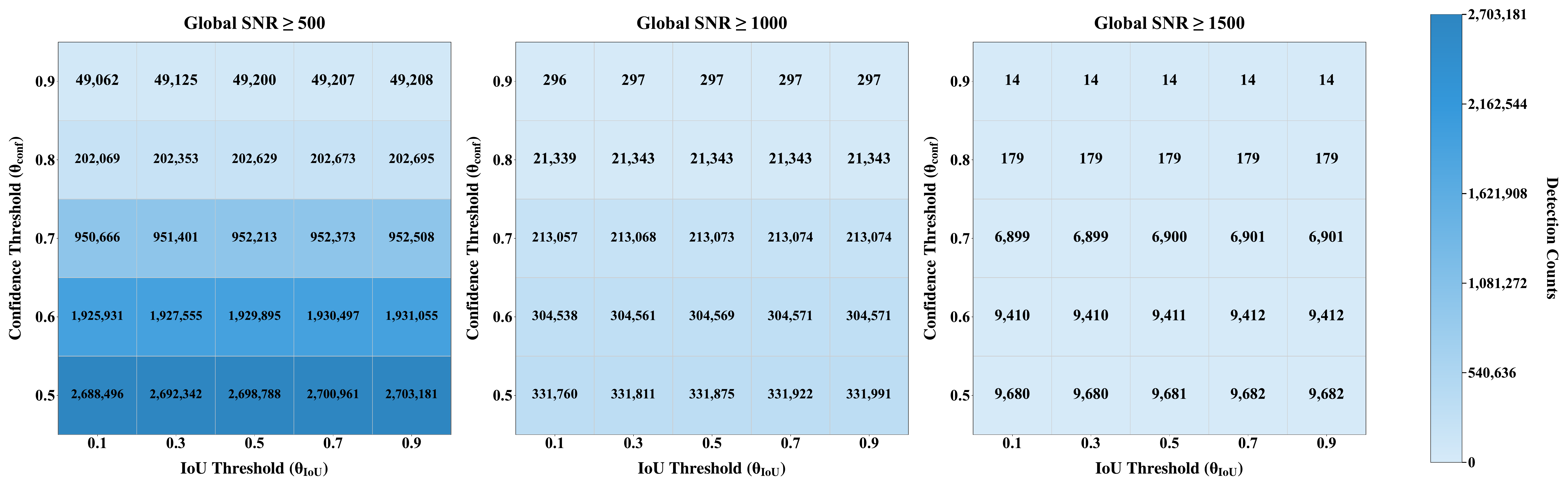}
\caption{Sensitivity of detection yield to post-processing thresholds. Each panel shows a heat map of the retained fraction of detections versus the confidence cutoff and the NMS IoU threshold, under different global-SNR activation settings.}
\label{fig:filtering}
\end{figure*}

Figure~\ref{fig:filtering} quantifies how the retained yield depends on three hard post-processing knobs: the global-SNR activation gate ($\tau_{\rm glob}$), the confidence cutoff ($\theta_{\rm conf}$), and the NMS IoU threshold ($\theta_{\rm IoU}$). As expected, the retained fraction varies monotonically with $\theta_{\rm conf}$ and $\tau_{\rm glob}$, while it is comparatively insensitive to $\theta_{\rm IoU}$: across the tested settings, changing $\theta_{\rm IoU}$ produces $<1\%$ variation in the retained fraction even in the worst case, indicating that the output volume is controlled primarily by confidence gating rather than overlap suppression. In practice, we select $\tau_{\rm glob}=1000$, $\theta_{\rm conf}=0.7$, and $\theta_{\rm IoU}=0.9$ by applying pipeline inference to synthetic injected data, under the constraint of achieving 90\% recall on the full prediction-level positive samples. This operating point favors controllable sensitivity while preserving high-quality outputs. A higher $\tau_{\rm glob}$ ensures that the retained yield is driven primarily by the core feature-extraction backbone (MSWNet), since this gate directly selects on the strength of the network’s signal response. The choice of $\theta_{\rm conf}=0.7$ is further constrained by the requirement of target completeness, while simultaneously suppressing contamination from spurious confidence activations on negative samples, thereby stabilizing downstream candidate statistics under high-throughput inference. Under this configuration, the pipeline produces $139{,}127$ \textit{detections} (i.e., post-NMS outputs at the single-beam scale), which aggregate into $6{,}402$ multi-beam \textit{events} for veto evaluation, and ultimately yield $803$ \textit{candidates} after the all-beam anticoincidence veto.

\begin{figure*}[htbp]
\centering
\includegraphics[width=\linewidth, trim=2cm 2.5cm 2cm 4.8cm, clip]{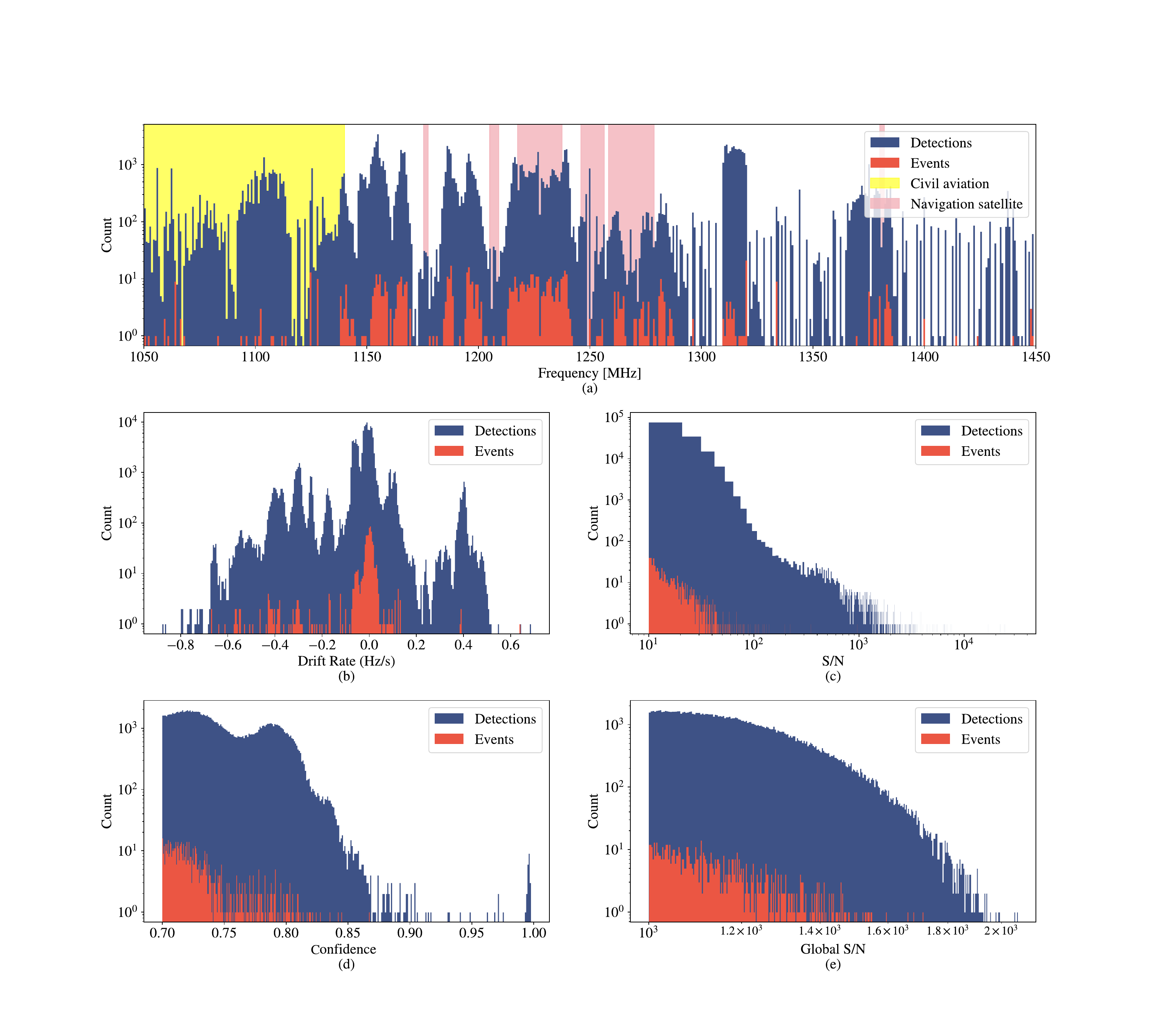}
\caption{Distributions of detected-signal parameters at different pipeline stages. Histograms show detections, events, and final candidates as functions of observing frequency, drift rate, S/N measured on raw data, global SNR on the cleaned map, and model confidence.}
\label{fig:distributions}
\end{figure*}

To characterize the population-level behavior of the search products, we summarize the distributions of key parameters at each stage (detections, events, candidates) in Figure~\ref{fig:distributions}.

The frequency histograms show candidates distributed across the full 1.05--1.45~GHz band, with prominent clustering at known interference sub-bands rather than any preference for special ``magic'' frequencies. The drift-rate distributions are strongly concentrated near $\dot{\nu}\approx 0$, with nonzero drifts dominated by small magnitudes and an asymmetric sign balance, as expected when the observed population is largely shaped by terrestrial transmitters and satellite-like emitters. The S/N distributions are bottom-heavy: most surviving products sit near the minimum S/N validation cutoff, with only a small tail of high-S/N cases. Finally, because we apply a strict confidence gate, the candidate confidence distribution is skewed toward high values, which is operationally favorable: the system tends to output either high-confidence signal-like tracks or nothing, reducing manual triage volume.

\subsection{Notable Events}

{We use visual inspection to identify the most plausible candidates based on the following criteria: (1) residual beam-to-beam anticoincidence checks on candidates passing the standard on–off veto pipeline, (2) exclusion of known RFI morphologies (e.g., harmonics and stationary instrumental artifacts), and (3) consistency of temporal stability and frequency drift with narrowband signal behavior. After this, only two notable narrowband events remain. The first is the previously reported signal toward Kepler-438 at 1140.604 MHz\citep{2022AJ....164..160T} (hereafter NBS 210629, where “NBS” means narrowband signal and “210629” is the date we detected it), which initially appeared consistent with an ETI technosignature but was ultimately excluded (it is discussed as a representative case in Section~\ref{sec:prior}). The other one is identified from the K2-155 observation. K2-155 is an early M-dwarf at $\sim$72.93 pc, known to host three transiting super-Earth planets\citep{2018AJ....155..124H}, including an outer planet near the habitable zone (K2-155d, $P\approx40.7$ days, $R\approx1.6~R_\oplus$). This system’s habitable-zone super-Earth made it an intriguing SETI target. The event (hereafter NBS 260108) is reported as a narrowband drifting signal at 1148.4167 MHz, and the signal exhibits a downward drift at approximately $-0.038$ Hz/s over the observation. Its total-intensity (Stokes~$I$) signal-to-noise ratio is $S/N\approx15$. Figure~\ref{fig:19beam} illustrates the dynamic spectrum across all beams. 

\begin{figure*}[htb]
\centering
\includegraphics[width=\textwidth]{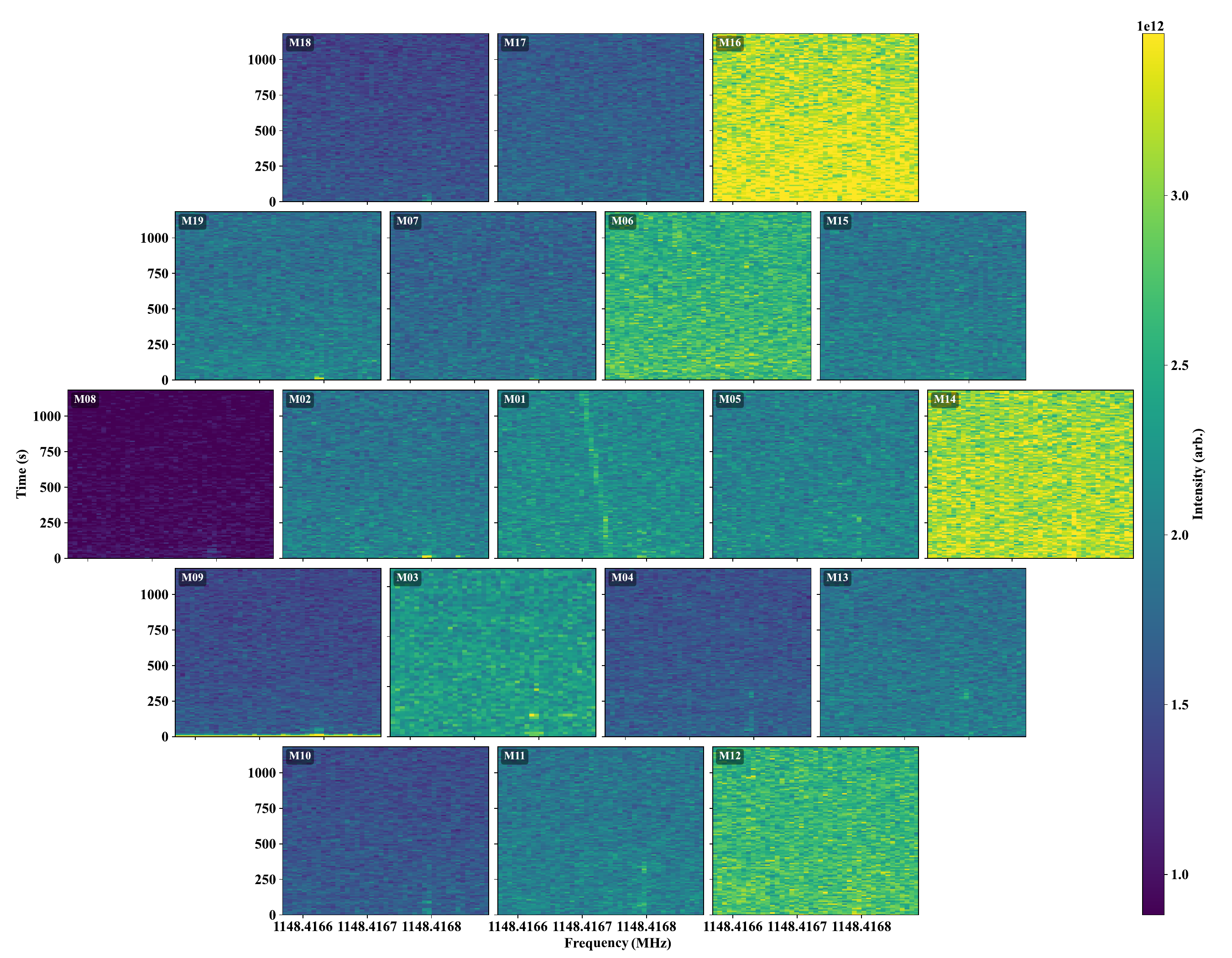}
\caption{Dynamic spectrum (frequency-time waterfall) from the 19-beam L-band receiver during the K2-155 observation, centered on the candidate frequency. The narrow drifting signal (1148.4167 MHz, drift $-0.038$ Hz/s) appears in Beam 1 (on target) and is not detected in any of the other 18 beams.}
\label{fig:19beam}
\end{figure*} 

The observed drift rate of $-0.038$ Hz/s implies a line-of-sight acceleration of order $a_{\text{LOS}}\sim (\dot \nu/\nu)c \approx (0.038~\text{Hz/s}/1148~\text{MHz})c \sim 10^{-2}$ m/s$^2$, which is physically plausible for Doppler drift from relative motion\citep{2022ApJ...938....1L}. Using $\dot{\nu} = \frac{\nu}{c}\frac{d v_{\rm LOS}}{dt}$, we estimate contributions include Earth’s rotation, Earth’s orbital motion, and potential transmitter motion. Earth’s diurnal rotation ($\omega_\oplus\approx7.29\times10^{-5}$~s$^{-1}$) yields a maximal radial acceleration at FAST’s latitude of order $\omega_\oplus^2 R_\oplus \cos25^\circ \sim3\times10^{-2}$ m/s$^2$, while Earth’s orbital motion contributes $\sim6\times10^{-3}$ m/s$^2$. If the transmitter were on an exoplanet, its orbital acceleration could be comparable or larger; e.g., K2-155d ($P\approx40.68$ d, $a=0.1886$ AU\citep{2018AJ....155..124H}) gives $a_p\approx4\pi^2 a/P^2\sim9\times10^{-2}$ m/s$^2$. Hence a net relative acceleration of $\sim10^{-2}$–$10^{-1}$ m/s$^2$ is expected, placing $|\dot \nu|\approx3.8\times10^{-2}$ Hz/s well within bounds for a genuine sky signal; the negative sign indicates increasing line-of-sight separation during the scan. We analyze this signal in the following section and elucidate its potential origin.

\section{Discussion and Conclusions}
\label{sec:discussion}

\subsection{Further Identification of NBS 260108}
\label{sec:further_identification}

\subsubsection{Polarization signature and cross-target recurrence}
\label{sec:nbs260108-selfcheck}

\begin{figure*}[htb]
\centering
\includegraphics[width=\linewidth]{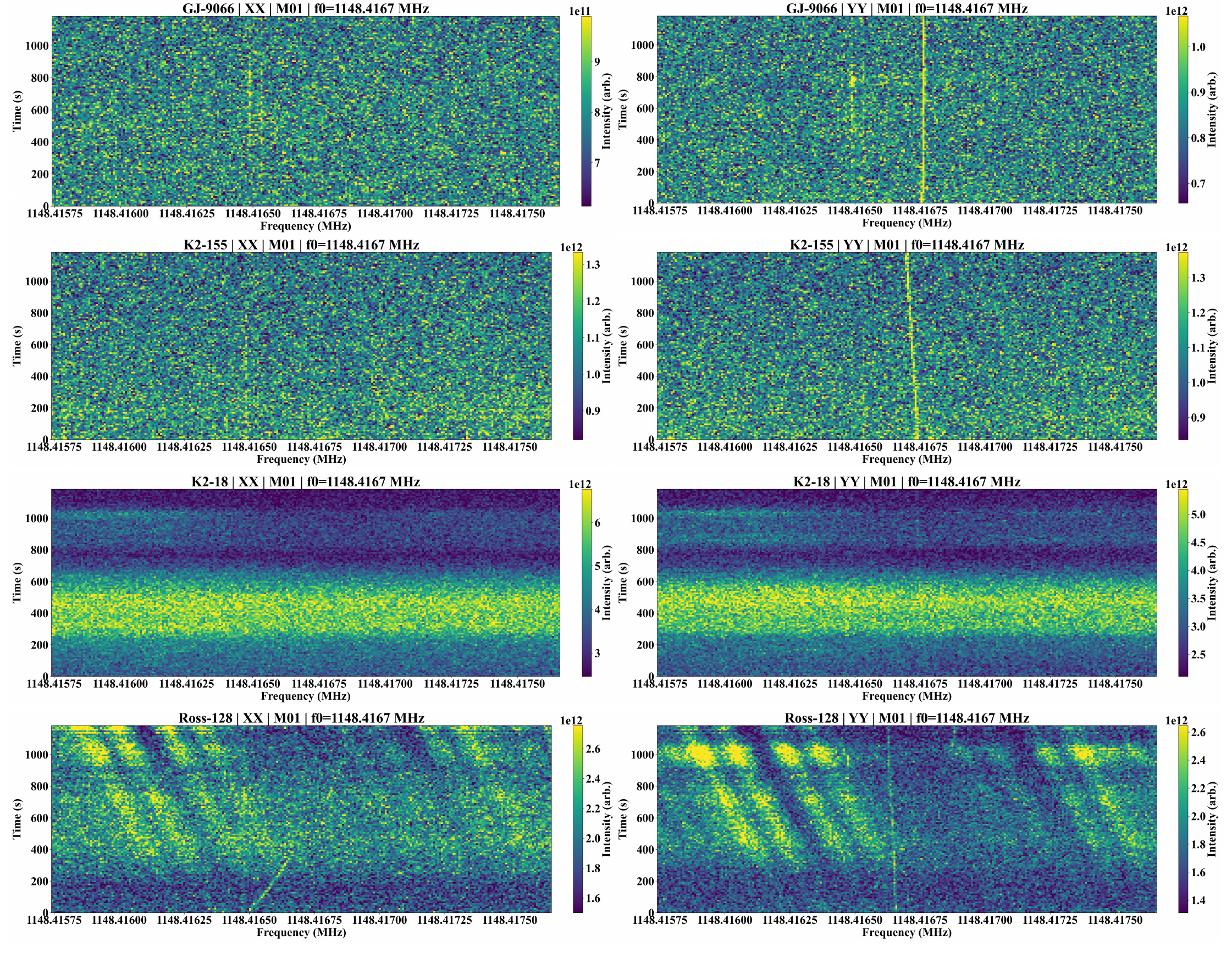}
\caption{Comparison of the 1148.4167 MHz region in XX vs. YY polarization for four targets observed on 2021-09-10 (arranged by time).}
\label{fig:obs_xx_yy}
\end{figure*}

NBS~260108 appears as a continuous, unmodulated narrowband tone with an approximately constant drift rate, and no comparable drifting tone was found at the same frequency in the K2-155 pointing. Inspection of the two linear polarization products shows that the candidate is strongly detected in $YY$ alone (S/N~$\approx23$) and is absent in $XX$, indicating a highly asymmetric feed-frame polarization signature. Figure~\ref{fig:obs_xx_yy} shows this contrast directly. Although polarization alone is not a decisive discriminator, a single-feed response is a common warning sign of instrumental or terrestrial contamination\citep{Tinbergen1996AstronomicalPP}. 

The measured drift rate, $-0.038~\mathrm{Hz/s}$, corresponds to a line-of-sight acceleration of order $10^{-2}~\mathrm{m/s^2}$ at 1148 MHz. This value is not by itself anomalous for a possible sky signal, so the rejection must depend on polarization, beam behavior, and recurrence rather than on drift rate alone. We therefore used the same-day multi-beam and polarization products as a cross-target RFI check. The K2-155 scan was part of a 2021 September 10 sequence that also included K2-18, GJ-9066, and Ross~128. (1) A similar signal appears only in the $YY$ polarization, with no corresponding detection in $XX$. (2) The signal is localized to a single beam (Beam 1 for the target pointing), with no significant signal in the other beams, consistent with a direction-specific source. (3) The signal matches the K2-155 candidate in morphology (a straight, narrowband drift) and frequency. YY-dominated near-frequency counterparts were found in all three comparison pointings. Their detailed properties differ from NBS~260108---for example, Ross~128 is offset by $\sim0.02$ MHz, K2-18 is very weak, and GJ-9066 is nearly zero-drift---but their recurrence at the same instrumental polarization and frequency neighborhood is inconsistent with a signal unique to K2-155. The Ross~128 and K2-18 features fall below the significance threshold of pipeline, while the GJ-9066 feature was rejected as RFI. Thus, despite differences in drift rate and significance, the repeated appearance of narrowband features dominated by $YY$ near the same frequency strongly favors a common interference origin over a signal located at K2-155. 

Figure~\ref{fig:obs_xx_yy} overlays the spectra for all four targets in both polarizations. An analogous cross-target polarization/beam comparison was used to identify the  candidate Kepler-438 as RFI \citep{2024JPhCS2877a2053Z}.

\subsubsection{External emitter-class screening}
\label{sec:nbs260108-rfi-external}
We examined plausible RFI explanations for NBS 260108 by scoring four common L-band emitter classes against two operational criteria: (1) whether the source class can produce a narrowband, slowly drifting, strongly linearly polarized track like the event, and (2) whether it can plausibly persist across multiple independent pointings within the same observing session. Table~\ref{tab:rfi} summarizes this screening with representative systems for each class\footnote{Data were compiled from official sources, including FAA/ICAO standards: \url{https://www.faa.gov/air_traffic/publications/atpubs/aim_html/chap4_section_5.html}; \url{https://www.icao.int/filebrowser/download/4371?fid=4371}; FCC/ITU filings: \url{https://docs.fcc.gov/public/attachments/fcc-20-48a1.pdf}; GPS/BeiDou ICDs: \url{https://www.gps.gov/interface-control-documents-icds-interface-specifications-iss}; \url{http://en.beidou.gov.cn/SYSTEMS/ICD/201806/P020180608518432765621.pdf}; NASA/NRAO reports: \url{https://science.nrao.edu/facilities/vlba/observing/rfi}; DoD MIL-STD-6016: \url{https://quicksearch.dla.mil/qsDocDetails.aspx?ident_number=123964}; CJCSI 6232, and NTIA spectrum management documents: \url{https://www.ntia.gov/report/2023/national-spectrum-strategy-pdf}.}.

\begin{splitdeluxetable*}{llccBl}
\tablecaption{Overview of plausible RFI source classes evaluated for the 1148 MHz candidate signal. Condition~1 assesses consistency with the observed narrowband morphology, slow drift rate ($\sim0.04~\mathrm{Hz/s}$), and strong linear polarization. Condition~2 evaluates whether the source could plausibly affect multiple independent target pointings within a single observing session.\label{tab:rfi}}
\tablehead{
\colhead{Source Type} &
\colhead{Coupling / Signature} &
\colhead{Cond. 1 Matches} &
\colhead{Cond. 2 Match} &
\colhead{Representative Known Cases}
}
\startdata
Aeronautical beacon (DME/TACAN) & Pulsed; sidelobe pickup; not tone-like & None/Low/Medium & Medium & DME/TACAN channel 61Y (ground reply 1148~MHz); SSR/Mode~S (1090~MHz downlink) \\
Satellite communication & Persistent illumination; leakage/spurs possible & Medium/Low/None & High & Ligado Networks (1164--1215~MHz); Inmarsat Aero-L (1090--1195~MHz) \\
GNSS spurious signals & Global carriers; spurs/intermod under nonlinearity & Medium/Low/None & High & GPS/Galileo L5/E5a (1176.45~MHz); BeiDou B2a (1176.45~MHz) \\
Military radar / communication & Hopping/pulsed; intermittent; sidelobe pickup & None/None/Medium & High & Link-16 (frequencies 1113--1206~MHz); AN/TPS-77 (960--1215~MHz) \\
\enddata
\tablecomments{For satellite communications, FCC filings for Ligado Networks and ITU-R/NRAO reports for Inmarsat Aero-L note that leakage/harmonics can generate spurious and intermodulation products near 1148~MHz. For GNSS, NASA reports and GNSS ICDs document that receiver nonlinearities or transmitter emissions can produce nearby spurs, harmonics, and intermodulation products from GNSS carriers.}
\end{splitdeluxetable*}

Aviation systems in the 960–1215~MHz region (e.g., SSR/Mode S and DME/TACAN) are predominantly short pulsed bursts or pulse pairs \citep{7119060} and therefore do not naturally reproduce a smooth, continuous drifting track, making them a weak match to Condition 1 despite being in-band. Military emitters such as Link-16 and many L-band radars are likewise typically pulsed, hopping, or broadband/chirped, so they are more consistent with intermittent contamination than with a stable tone-like drift repeating across pointings, unless atypical operating modes are invoked. In contrast, satellite and GNSS illumination is globally persistent \citep{2022AJ....164..160T,2023AJ....165..132L}, can enter the telescope through sidelobes \citep{2021RAA....21...18W}, and can generate narrow intermodulation products displaced from the parent bands through receiver/backend nonlinearity driven by strong external fields \citep{ECC2011IridiumRFI}. However, they face notable difficulties: their known carrier frequencies are separated from 1148~MHz by tens of MHz, requiring specific and unverified intermodulation pathways; moreover, they do not naturally account for the high degree of linear polarization observed. Consequently, the candidate remains unclassified: it is not confirmed as a technosignature, and a definitive identification of an external contaminating source requires additional observations capable of establishing a direct emitter association. Note that we cannot determine from these data alone whether the apparent repetition across pointings reflects a true physical association or a chance coincidence; we therefore treat Condition~2 as a working assumption rather than a verifiable fact.

\subsubsection{Systematics and ensemble diagnostics}
\label{sec:nbs260108-systematics}

\begin{figure*}[t]
\centering
\includegraphics[width=\linewidth]{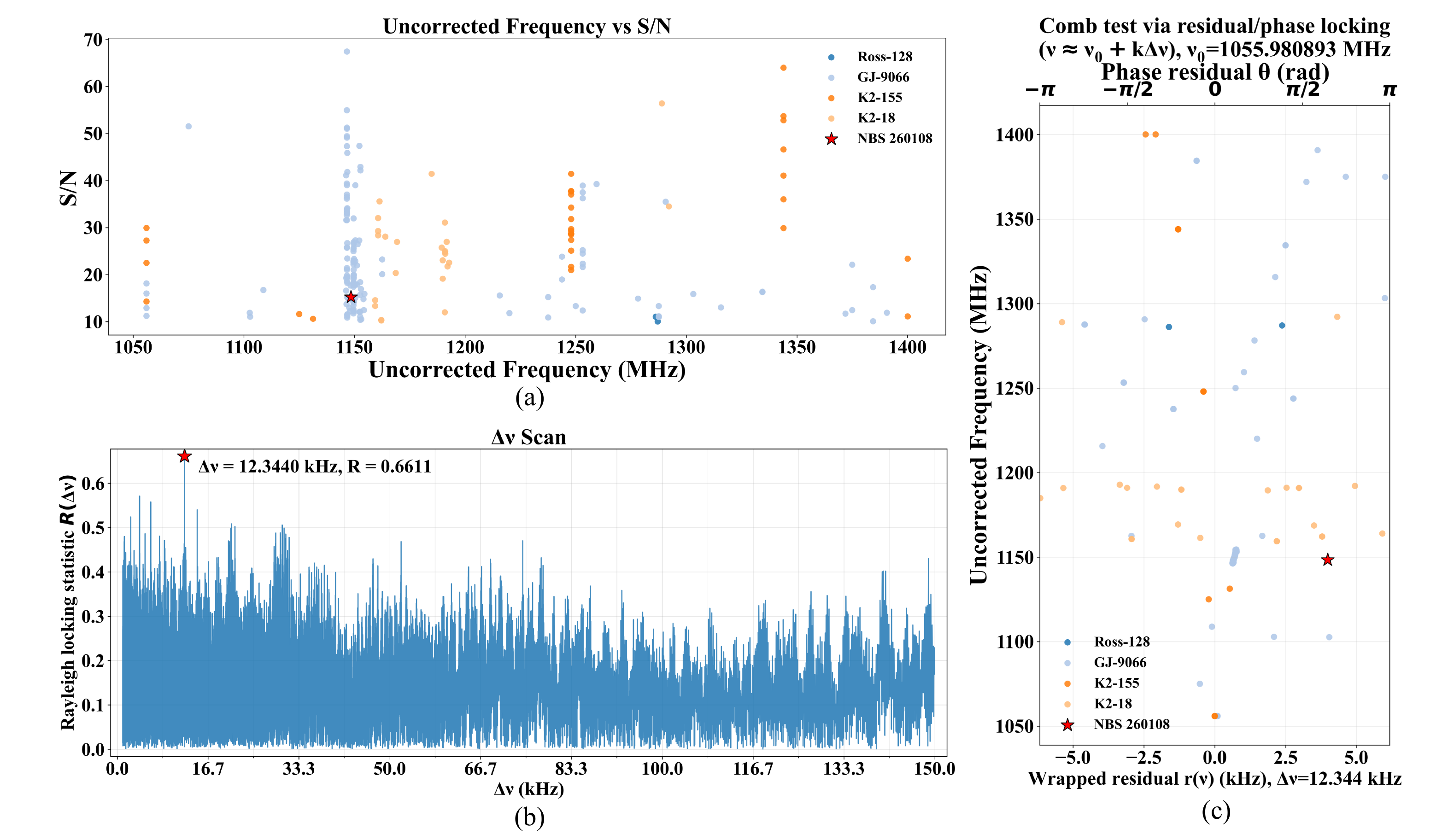}
\caption{Ensemble diagnostics for the matched-control hits. (a) Frequency--S/N distribution of the 201 matched hits, colored by group. (b) Rayleigh-style spacing scan, with the strongest response at $\Delta\nu^*\simeq12.344~\mathrm{kHz}$. (c) Wrapped residual/phase structure computed at $\Delta\nu^*$ with $f_0=\min(f_i)$, showing that the locked band extends across the survey frequency range.}
\label{fig:diagnostics}
\end{figure*}

Next, we examine whether NBS~260108 is embedded in a larger population of same-session systematics. We constructed a matched-control set using the same search configuration, restricted to the FAST L-band range (1.05--1.45~GHz) and to the drift-rate corridor of the K2-155 event, $-0.038024 \pm 0.006297~\mathrm{Hz/s}$. This selection produced 201 hits, including 31 from the K2-155 pointing. Within the K2-155 subset, all hits except NBS~260108 lack the same Beam-1-only localization pattern. Thus, the candidate's beam behavior is not the default outcome of this matched search corridor.

The matched hits are not randomly distributed in frequency. Instead, they cluster into discrete frequency groups, as shown in Figure~\ref{fig:diagnostics}(a), indicating a structured interference or instrumental environment. To test whether this population contains a preferred spacing, we applied a Rayleigh-style phase-locking scan. For each trial spacing $\Delta\nu$, we computed
\begin{equation}
R(\Delta \nu)=\left|\frac{1}{N}\sum_{i=1}^{N}\exp\left(2\pi j\nu_i/\Delta\nu\right)\right| .
\end{equation}
The strongest response occurs at $\Delta \nu^*=\arg\max_{\Delta \nu} R(\Delta \nu)\ \approx\ 12.3\ \mathrm{kHz}$ (Figure~\ref{fig:diagnostics}(b)), suggesting a dominant interval in the matched-hit ensemble. Because a peak in $R(\Delta\nu)$ can also be produced by a localized sub-band artifact, we further examined the coherence of this spacing across the full band using a residual-phase representation. We adopted $f\simeq f_0+k\Delta\nu$ with $f_0=\min(f_i)$ as a fixed phase reference and, for each hit, computed the wrapped residual $r_i=\mathrm{wrap}(f_i-f_0,\Delta\nu)\in[-\Delta\nu/2,\Delta\nu/2)$ and the corresponding phase $\theta_i=2\pi r_i/\Delta\nu\in[-\pi,\pi)$. If the 12.344 kHz peak were driven only by a narrow local feature, the residuals would align only within that frequency range. Instead, the residual band remains coherent across much of the 1.05--1.45~GHz window (Figure~\ref{fig:diagnostics}(c)). This behavior upgrades the $\Delta\nu$ peak from a numerical maximum in the scan to evidence of a survey-wide periodic-spacing component in the matched-hit population.

We also compared the matched hits with known Roach2 backend harmonic loci. Narrowband interference from Roach2 FPGA backends has been reported at frequencies expressible as integer combinations of the 33.3333 and 125.00~MHz oscillator clocks, corresponding to an oscillator-lattice step of $8.333333\ldots\,\mathrm{MHz}$ \citep{2022AJ....164..160T}. In our matched sample, the representative Roach2 loci at 1066.6658, 1124.9967, 1200.0104, and 1333.3270~MHz have zero matches. The 1375.0009~MHz locus has two GJ-9066 matches, and the 1400.0207~MHz locus has two K2-155 near-matches; the latter exceeds the nominal oscillator-drift tolerance and is therefore only suggestive. This comparison does not uniquely identify the matched population with a Roach2 oscillator lattice, but it keeps backend-related structure within the plausible RFI space.

Under the fixed spacing $\Delta\nu^*$, the 30 non-candidate K2-155 hits remain strongly locked to the same residual band as the wider matched population ($R_{\mathrm{K2,bulk}}=0.9483$). Including NBS~260108 lowers the K2-155 locking strength to $R_{\mathrm{K2,all}}=0.8954$. Together with its Beam-1-only localization, this places NBS~260108 outside the dominant $\sim12.3$~kHz periodic-spacing family, rather than making it a typical member of that interference-driven population. A complementary morphology/beam montage is retained in the supplementary material.

Finally, we inspected the local spectral neighborhood of NBS~260108. A faint comb of narrow carriers is present near 1148~MHz with a spacing of $\sim1$~kHz; one comb line lies $\sim250$~Hz below the candidate frequency. These carriers are frequency-stable (drift $<10^{-4}~\mathrm{Hz/s}$, consistent with zero within measurement error), appear in multiple beams, and are therefore rejected by the single-beam localization veto. Similar $\sim1$~kHz combs occur in 17 of the 33 stellar observations, implying a common RFI component. Because this local comb is temporally variable and its coupling mechanism is not identified, it should not be treated as a unique explanation for NBS~260108. However, it does strengthen the case that the candidate lies in a structured interference environment.

Taken together, the ensemble diagnostics narrow but do not close the interpretation. NBS~260108 is not a typical member of the dominant $\sim12.3$~kHz periodic-spacing population, yet the same data show a structured RFI landscape that includes local $\sim1$~kHz spur trains, possible backend-related structure, extreme $YY$-only response, and near-frequency recurrence across same-session pointings. In combination with the external emitter screening above, these factors make an anthropogenic or instrumental origin more plausible than an extraterrestrial one. We therefore treat NBS~260108 as a low-priority technosignature candidate and not as a robust detection. Its specific coupling mechanism remains unresolved, so it is retained only as an internal follow-up item for future observations with stronger cross-target, beam, and polarization discrimination.

\subsubsection{Comparison with prior analyses of the same FAST observing campaign}
\label{sec:prior}

Our target list and observing configuration overlap those of \citet{2022AJ....164..160T} and \citet{2023AJ....165..132L}, which used FAST multi-beam coincidence matching as a primary RFI-rejection framework. We adopt the same anticoincidence principle \citep{2005RaSc...40.5S18H}, but apply a stricter downstream veto by treating all non-primary beams as OFF-source channels and rejecting events with matched OFF-beam counterparts within $\pm 3\,\delta\nu$. In parallel, for the main pipeline we form total intensity ($XX+YY$) for first-pass processing to stabilize feature extraction, while retaining XX/YY for downstream diagnostics. This design reflects a deliberate cost–benefit trade for measurability and comparability: first-pass narrowband searches are typically performed on Stokes-$I$, while polarization products are more valuable as a diagnostic layer than a hard detection dependency. For example, \citet{2021NatAs...5.1148S} run their narrowband search solely on Stokes-$I$ and do not use their commensal full-Stokes product because it was configured for flare science (high temporal resolution but coarse frequency resolution). Polarization-based criteria can be powerful for ETI–RFI discrimination but often require additional observing/analysis overhead \citep{2024AJ....167....8L}; we therefore reserve XX/YY behavior for downstream diagnostics and classification rather than expanding the first-pass scope.

\begin{deluxetable*}{llccccccc}
\tablewidth{0pt}
\tablecaption{Two representative FAST events from prior analyses and their recovery by our pipeline. Literature values are taken from \citet{2022AJ....164..160T} (Kepler-438; NBS~210629) and \citet{2023AJ....165..132L} (HD~180617; NBS~210421). \label{tab:prior_cases}}
\tablehead{
\colhead{Event} &
\colhead{Pol.} &
\colhead{$\nu_{\mathrm{uncorr}}$ (MHz)} &
\colhead{$\dot{\nu}$ ($\mathrm{Hz/s}$)} &
\colhead{S/N} &
\colhead{Beam} &
\colhead{Morphology} &
\colhead{Conf.} &
\colhead{${\rm SNR}_{\rm glob}$}
}
\startdata
Kepler-438 (NBS~210629) & XX & 1140.604 & $-0.0678$ & 9.6 & Beam~1 & -- & -- & -- \\
Kepler-438 (NBS~210629) & YY & 1140.604 & $-0.0678$ & 22.6 & Beam~1 & -- & -- & -- \\
Kepler-438 (this work) & XX & 1140.604012 & $-0.0772$ & 4.4 & Beam~1 & linear & 0.78 & 659.6 \\
Kepler-438 (this work) & YY & 1140.604010 & $-0.0713$ & 23.2 & Beam~1 & linear & 0.79 & 1116.6 \\
Kepler-438 (this work) & $\mathrm{XX}+\mathrm{YY}$ & 1140.604011 & $-0.0726$ & 19.9 & Beam~1 & linear & 0.80 & 1167.5 \\
\hline
HD~180617 (NBS~210421) & XX & 1404.050 & $-0.096$ & $\sim$5 & Beam~4 & -- & -- & -- \\
HD~180617 (NBS~210421) & YY & 1404.050 & $-0.096$ & $\sim$5 & Beam~4 & -- & -- & -- \\
HD~180617 (this work) & XX & 1404.050065 & $-0.1241$ & 7.8 & Beam~4 & linear & 0.76 & 1175.6 \\
HD~180617 (this work) & YY & 1404.050071 & $-0.1266$ & 11.1 & Beam~4 & linear & 0.77 & 1269.2 \\
HD~180617 (this work) & $\mathrm{XX}+\mathrm{YY}$ & 1404.050075 & $-0.1326$ & 14.7 & Beam~4 & linear & 0.80 & 1401.5 \\
\enddata
\tablecomments{$\nu_{\mathrm{uncorr}}$ is the uncorrected (observed) frequency. ``Pol.'' denotes polarization (XX, YY) or total intensity ($\mathrm{XX}+\mathrm{YY}$). ``Conf.'' is the estimator output score after post-processing. ${\rm SNR}_{\rm glob}$ denotes the global SNR statistic defined in Section~\ref{sec:pipeline}.}
\end{deluxetable*}

To connect our pipeline output directly with prior FAST reports, we use two published events as anchored recovery tests: the Kepler-438 event (NBS~210629) discussed by \citet{2022AJ....164..160T}, and the HD~180617 event (NBS~210421) reported by \citet{2023AJ....165..132L}. For both cases, we also run single-polarization inference in XX and YY. The recovered frequencies and drift rates are consistent with the published values at the level expected from independent patch extraction and post-processing. The Kepler-438 event retains the strong XX/YY imbalance reported previously, while the HD~180617 event remains weak and shows behavior consistent with an instrumental or beam-pair origin. Table~\ref{tab:prior_cases} summarizes the comparison.

Both published events are recovered by our pipeline before downstream screening, demonstrating sensitivity to the same class of narrowband drifting features identified in earlier FAST analyses. However, these cases also illustrate why single-beam localization and approximately linear drift are insufficient as standalone evidence for an astrophysical origin. Polarization asymmetry, OFF-beam recurrence, and beam-pair structure remain essential diagnostic layers, and they motivate the multi-stage verification applied to NBS~260108.

\subsection{Pipeline Interpretation and Comparison}
\label{sec:discussion_pipeline}

\subsubsection{Wavelet-Path Contribution}
\label{sec:wavelet}

Wavelet-based downsampling provides a direct basis for the low-frequency backbone of MSWNet. Prior wavelet-pooling work has shown that the LL approximation component can replace conventional pooling as an effective compact downsampled representation in convolutional networks \citep{williams2018wavelet}. Related wavelet-integrated CNNs further demonstrate that propagating DWT-derived low-frequency features through the network can improve structural preservation and noise robustness compared with standard downsampling operations \citep{9508165}. Following this, MSWNet uses the LL stream as a stable low-frequency propagation path for coherent track morphology, while retaining encoder-side high-frequency transfers to support fine-scale structural recovery.  

To assess the role of the detail pathway, we performed an inference-time ablation: all encoder-to-decoder high-frequency transfers were set to zero, forcing the reconstruction to rely on the low-frequency pathway alone. We find that in low-S/N regimes, particularly near the detection limit, removing the high-frequency pathway can substantially weaken the narrowband response and push it below detectability. This is important because the detection-limit regime defines the practical scientific reach of the search pipeline. The high-frequency pathway therefore contributes not only to reconstruction fidelity, but also to maintaining weak-signal evidence at the boundary of detectability. This point is also reflected in our candidate set: many of the 803 detections are located around ${\rm S/N}\sim10$, which is commonly treated as the practical detection limit. MSWNet still produces strong responses for these low-S/N candidates, with only minor qualitative differences from high-S/N cases.

This suggests that the low-frequency pathway carries the coarse track morphology, whereas the reused high-frequency pathway preserves fine-scale structure that becomes important for marginal signals and structured contamination. During training, we also observed negative bands in some denoised maps; under the wavelet synthesis view, these responses can be interpreted as cancellation-like terms associated with suppressing reused high-frequency structure rather than as detections themselves. Such internal negative responses could be thresholded or morphologically filtered to form a dense marker set for RFI-like structures rejected by the reconstruction stage, which we leave as a direction for future work.

\begin{figure}[htb!]
\centering
\includegraphics[width=240pt]{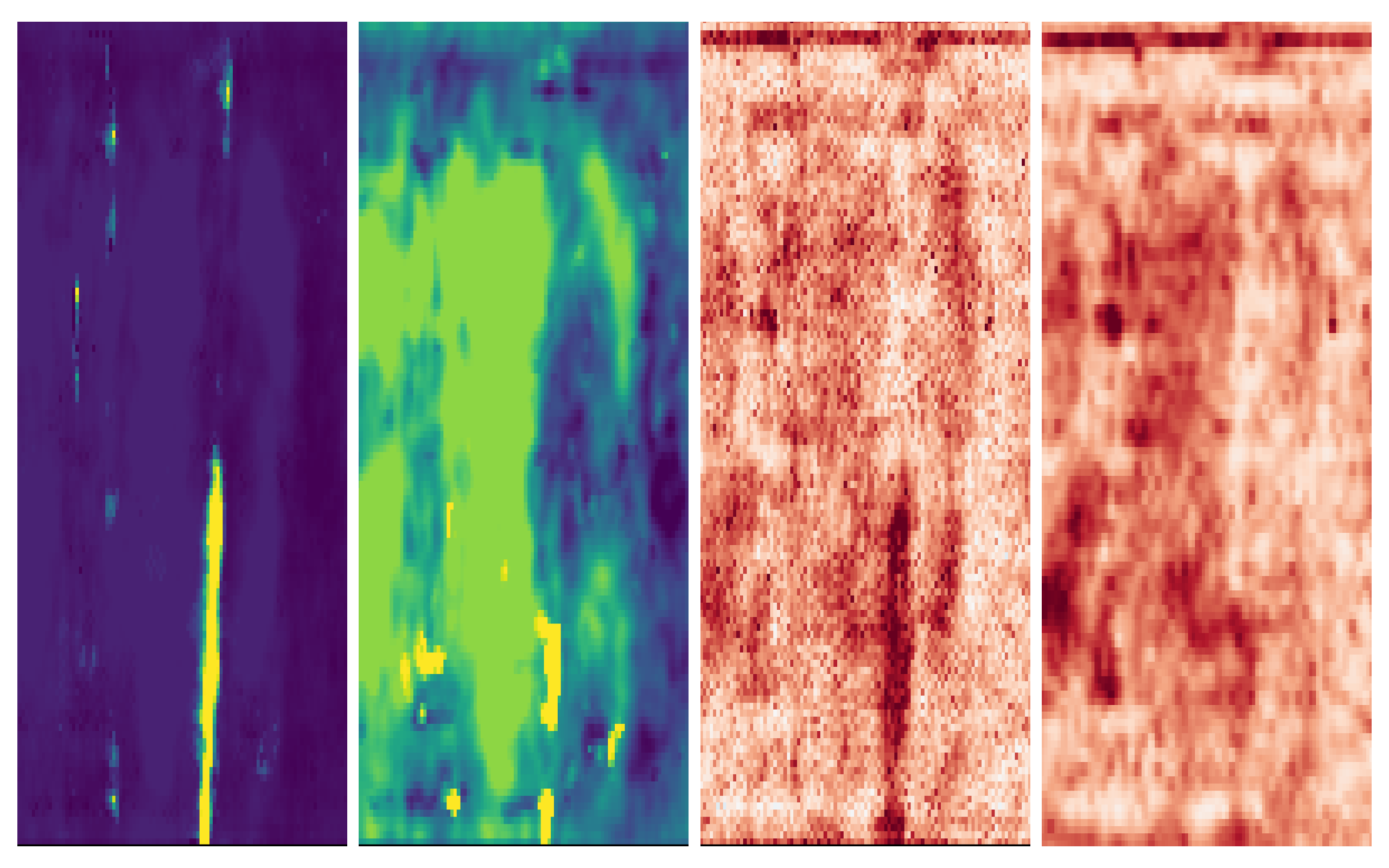}
\caption{Inference-time wavelet-path ablation. From left to right: denoised output, LL-only output, inverse-synthesized feature, and its LL-only counterpart.}
\label{fig:wavelet_ablation}
\end{figure}

\subsubsection{Methodological Comparison with Incoherent De-Doppler Drift Search}
\label{sec:comparison}

We compare our pipeline against the widely used incoherent de-Doppler drift search paradigm (Taylor-tree implementation; commonly deployed via \texttt{TurboSETI}), which searches for narrowband signals by summing power along linear frequency--time tracks over a discrete drift-rate grid \citep{2017ApJ...849..104E,2020AJ....159...86P}. At its core, this paradigm performs a matched-filter-like incoherent de-Doppler search over a discrete drift-rate grid. For each trial drift $\dot{\nu}$, it integrates power along a linear trajectory in the time--frequency plane,
\begin{equation}
P(\nu_0;\dot{\nu}) \propto \sum_{t} S\!\left(t,\,\nu_0+\dot{\nu}t\right),
\end{equation}
and promotes threshold crossings of $P$ (after normalization) to ``hits'' \citep{2017ApJ...849..104E}. This design is transparent and physically motivated for strictly linear drifts, but it couples the search complexity and output volume directly to the size of the drift grid and the local RFI environment. Under RFI-crowded conditions or when signals deviate from the simplest narrowband-tone assumptions, this coupling can materially impact practical efficiency and downstream hit volume \citep{2020AJ....159...86P,2021AJ....161..286T}.

The MSWNet pipeline replaces explicit drift-grid scanning with a two-stage ``feature extraction $\rightarrow$ parameter estimation'' formulation. MSWNet first produces a wavelet-consistent cleaned representation that suppresses broadband/impulsive structures while enhancing narrow drifting tracks across multiple scales. A lightweight estimator then regresses a compact set of predictions with $(f_{\mathrm{start}}, f_{\mathrm{stop}})$ and a confidence score, turning detection into low-cost parameter inference rather than enumerating all $\dot{\nu}$ hypotheses. Practically, this moves the decision boundary from a single hard S/N threshold to a staged, auditable sequence: (1) a robust activation gate on the cleaned map (global SNR); (2) confidence gating and IoU-based suppression (NMS) to convert predictions into single-beam detections; and (3) classical per-track S/N validation on the raw spectrogram to promote detections into veto-ready events \citep{2017ApJ...849..104E}.

A practical nuance of the \texttt{TurboSETI} workflow is the (sometimes enabled) “top-hit” reporting mode, in which only the highest-S/N hit is kept within a de-duplication window around a given spectral neighborhood. Because that window effectively scales with both the observation duration and the maximum drift-rate setting, increasing ${\rm MDR}$ can counterintuitively decrease the number of reported hits by merging multiple nearby threshold crossings into a smaller set of top-ranked outputs. This is therefore an output-reporting artifact, not a genuine sensitivity improvement. Our pipeline does not show this behavior: it produces a fixed number of predictions per patch and then applies explicit, auditable post-processing (global SNR gating, confidence filtering, and IoU-based suppression), preserving the expected monotonic dependence of retained yield on thresholds (Figure~\ref{fig:filtering}).

These structural differences produce distinct failure modes that matter operationally in RFI-crowded data. In classical incoherent de-Doppler (Taylor-tree) searches, at least four false-positive phenotypes are commonly observed: (1) broadband impulsive interference can seed many drift trials and adjacent-frequency threshold crossings, creating dense hit clouds that require heavy downstream clustering; (2) high-curvature or oscillatory tracks (including quasi-sinusoidal structures from certain satellite modulations) are not matched by linear drift templates, yet can still produce fragmented threshold crossings and biased or even sign-flipped drift estimates when the algorithm latches onto local segments; (3) no-signal false hits can arise from stochastic maxima in the integrated power statistic over large drift grids; and (4) spotty/comb-like RFI (e.g., bursty carriers or periodically spaced sub-tones) can generate apparently narrowband hits that survive simple  S/N cuts but fail physical plausibility on inspection
\citep{2017ApJ...849..104E,2020AJ....159...86P}.

\begin{figure*}[htb!]
\centering
\parbox[t]{0.245\linewidth}{\centering
\includegraphics[width=\linewidth]{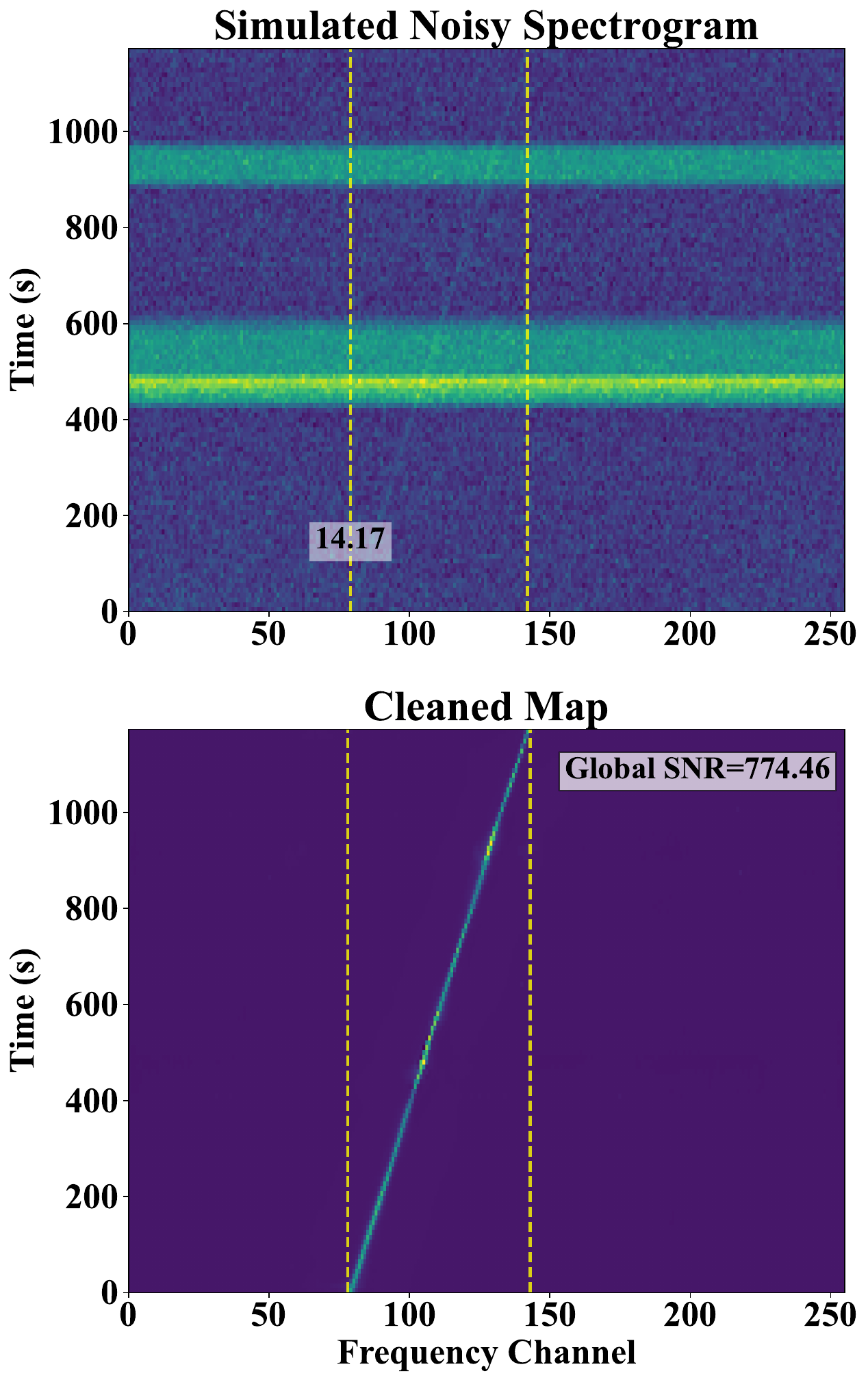}\\
\hspace{13pt}(a)}
\parbox[t]{0.245\linewidth}{\centering
\includegraphics[width=\linewidth]{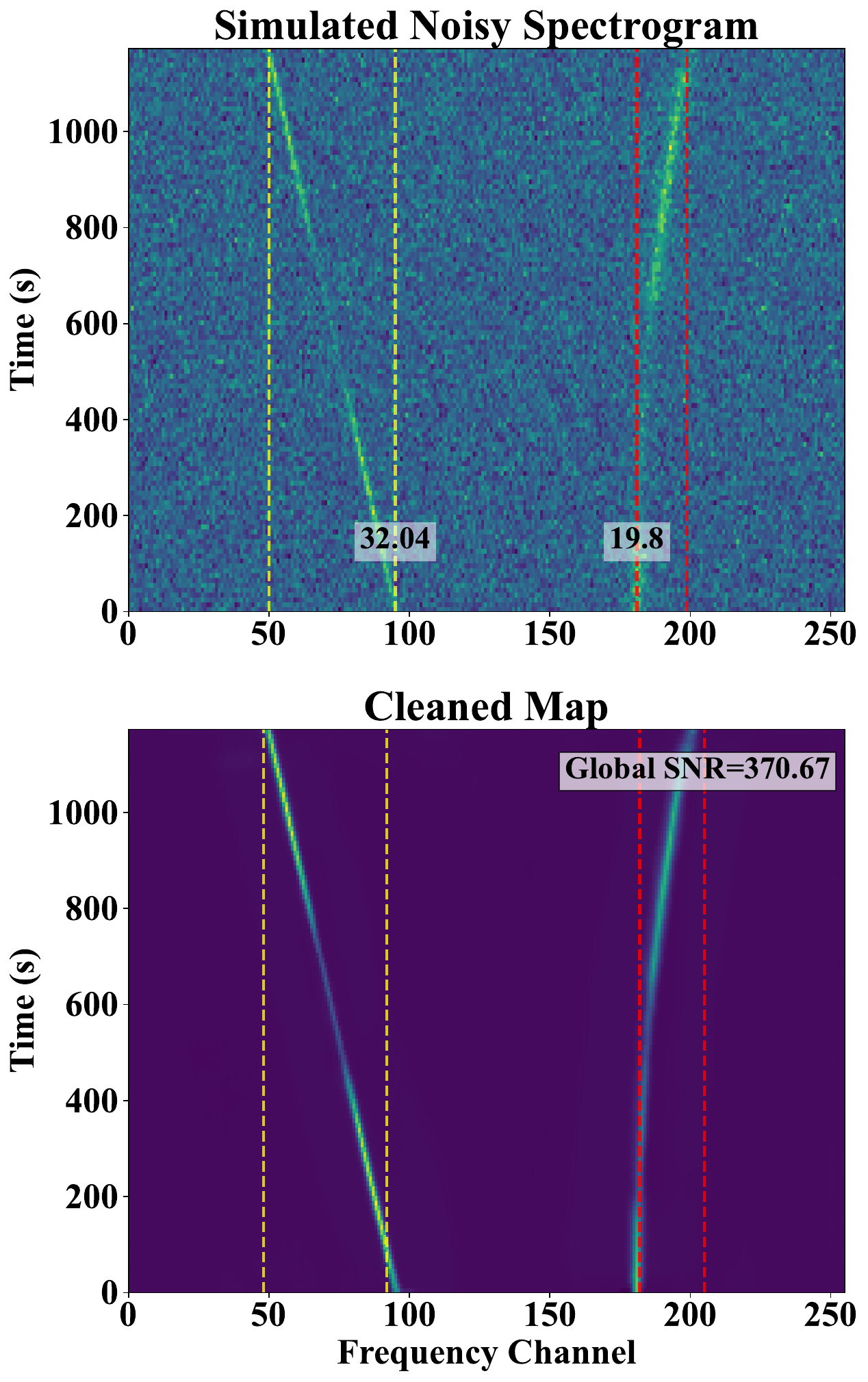}\\
\hspace{13pt}(b)}
\parbox[t]{0.245\linewidth}{\centering
\includegraphics[width=\linewidth]{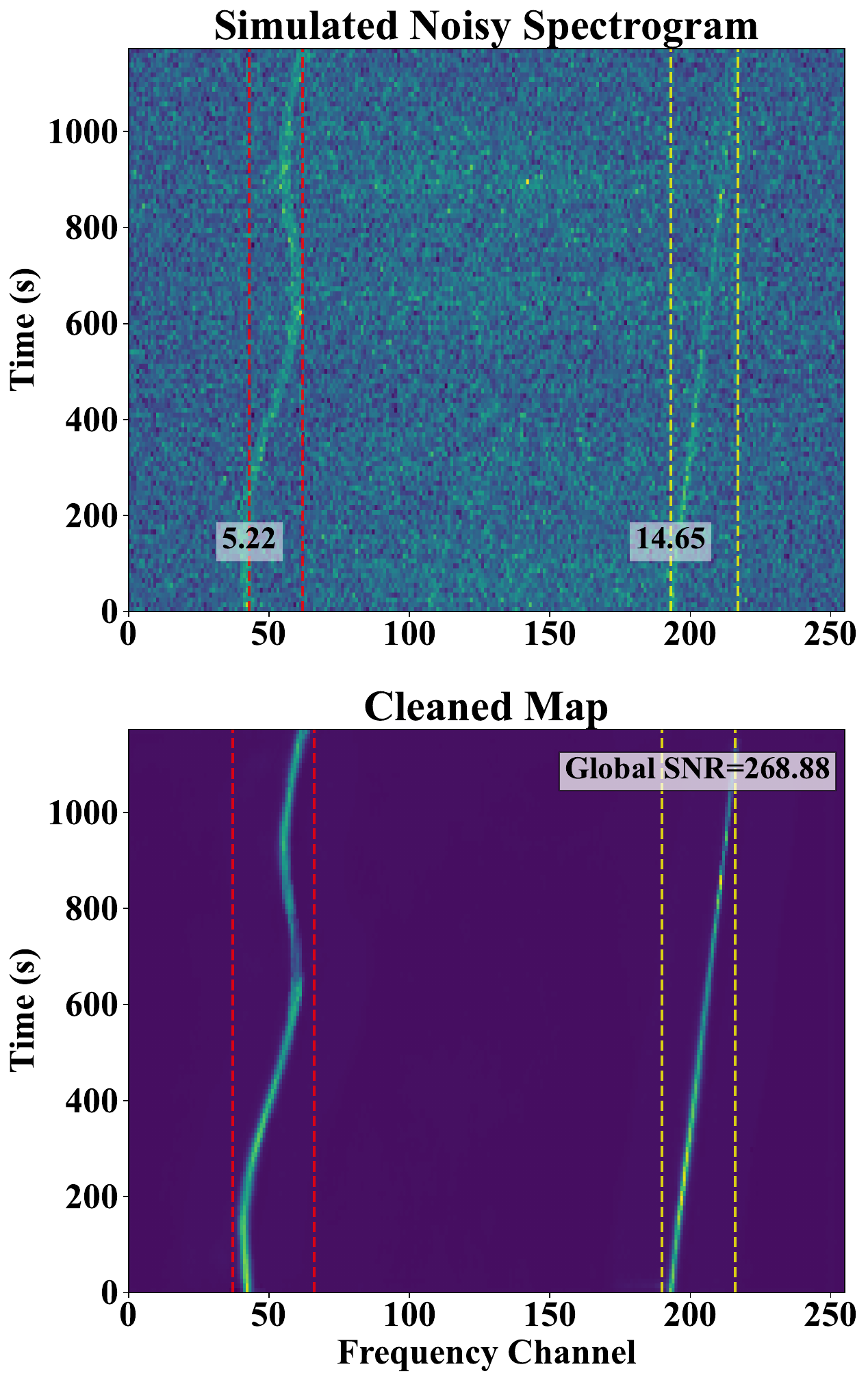}\\
\hspace{13pt}(c)}
\parbox[t]{0.245\linewidth}{\centering
\includegraphics[width=\linewidth]{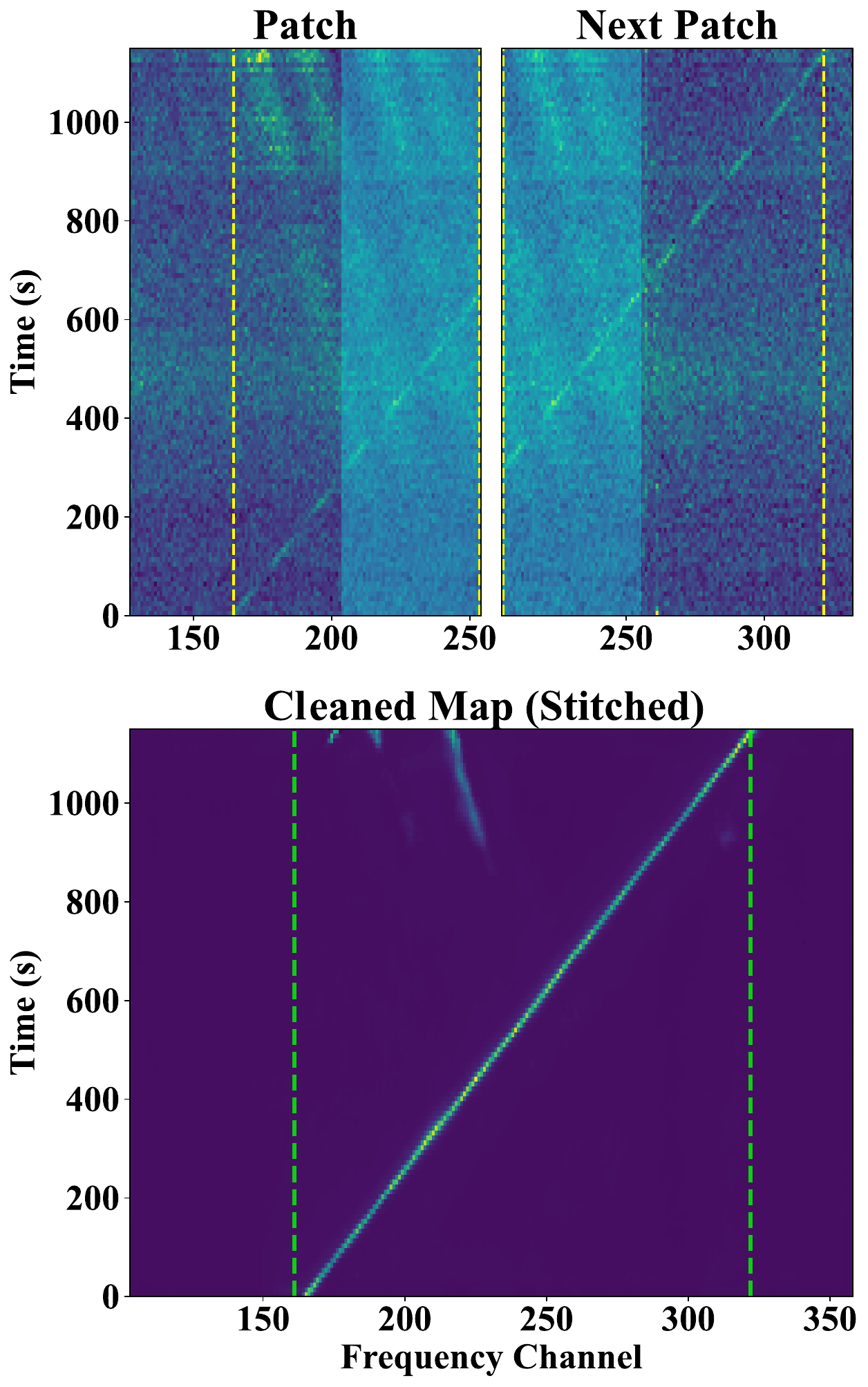}\\
\hspace{13pt}(d)}
\caption{Robustness gallery (left to right). Yellow boxes denote detections morphologically classified as linear, while red boxes indicate signals with curvature. (a) Target drifting narrowband track + broadband impulsive transient, with the corresponding MSWNet cleaned map and predicted box. (b) Target + curved interfering track, with the cleaned map and predicted box. (c) Target + quasi-sinusoidally modulated interference, with the cleaned map and predicted box. (d) Target spanning patch boundaries (tiling), with the stitching result across patches. The brightened area is the overlapping region and the green box is the equivalent detection obtained after cross-patch stitching.}
\label{fig:robustness_case}
\end{figure*}

Our approach is explicitly designed to reject most of these patterns earlier in the stack. MSWNet’s wavelet-constrained reconstruction preferentially preserves coherent, track-like morphology while attenuating broadband transients and incoherent bursts; this reduces the propensity for impulsive wideband interference to propagate into large hit clouds. Likewise, nonlinear structures are handled selectively: curved tracks consistent with our simulated curvature regime can be retained and \emph{explicitly flagged} as morphology mismatches for downstream categorization, whereas other atypical nonlinear patterns are suppressed early and do not survive into dense hit lists, which improves separability from truly linear drift candidates. Consequently, many failure cases that would inflate \texttt{TurboSETI} hit lists are filtered before the system ever produces detections, keeping the downstream event list compact even prior to multi-beam vetoing. Figure~\ref{fig:robustness_case} provides concrete visual evidence of these advantages across representative interference mixtures.

\begin{figure*}[htb!]
\centering
\includegraphics[width=\linewidth]{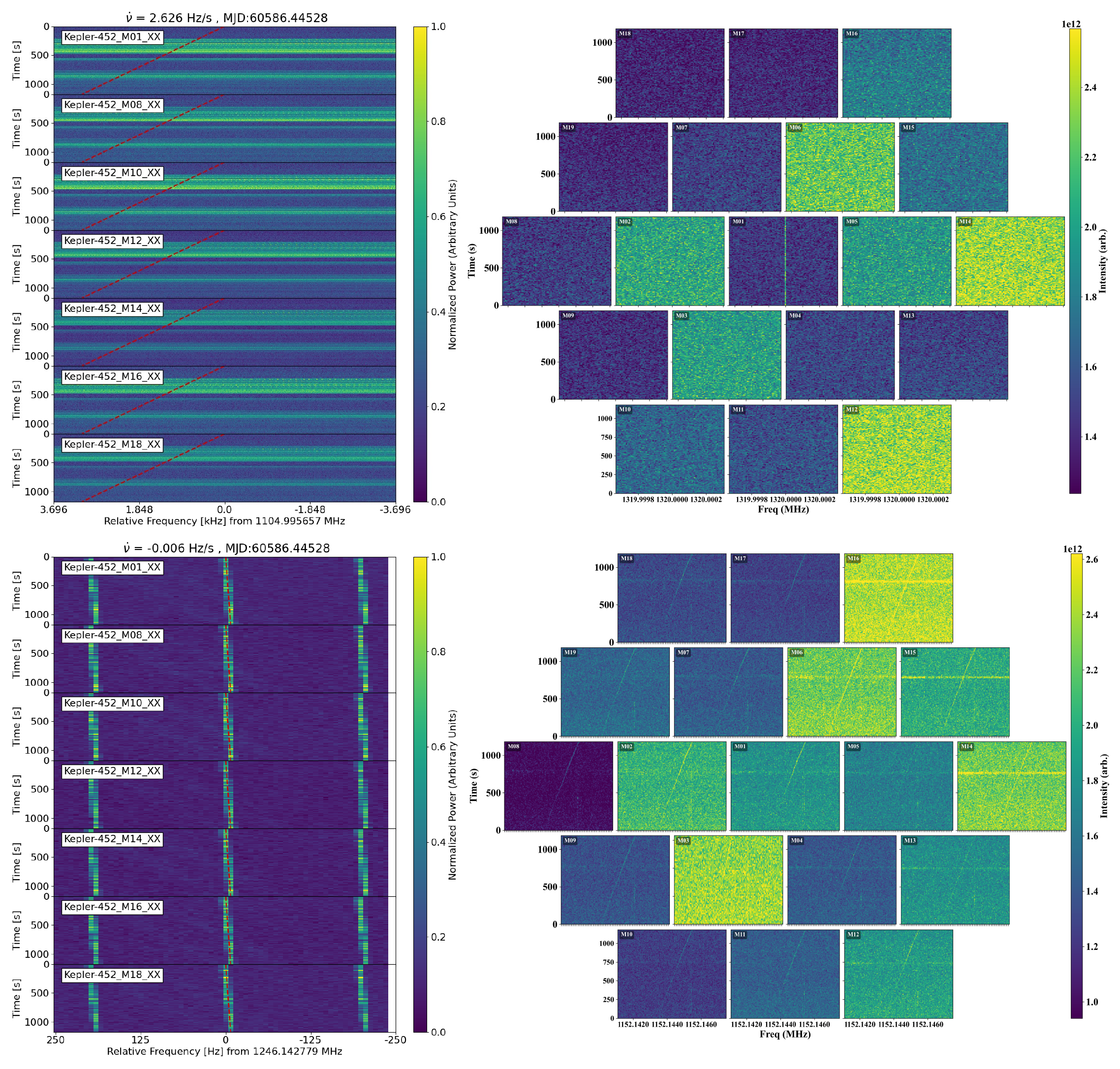}
\caption{Representative false positives for the incoherent de-Doppler drift search (left; Taylor-tree implementation in \texttt{TurboSETI}) and the MSWNet pipeline (right).}
\label{fig:false_positive}
\end{figure*}

This robustness is not free: compared to deterministic drift-grid scoring, the estimator’s frequency endpoint regression can exhibit small localization errors even when a signal is visually obvious. In an all-beam veto, this matters because coincidence matching is performed within a finite tolerance (e.g., $\pm 3\,\delta\nu$). Small per-beam localization biases can therefore increase the incidence of ``near-miss'' coincidences that evade strict matching, yielding more veto-failure survivors at the same tolerance than a purely template-based frequency estimator would. In our pipeline this is mitigated by (1) enforcing conservative post-processing (high confidence), (2) measuring final S/N on raw data with off-track background, and (3) performing veto at the event level under full 19-beam context. Representative failure modes for both approaches are summarized in Figure~\ref{fig:false_positive}.

We emphasize that this effect is a downstream localization error of the lightweight estimator rather than a limitation of MSWNet’s upstream feature extraction; in future work, we will explore replacing (or augmenting) the estimator with physics-driven track fitting (e.g., Hough-transform-based line/curve detection) to achieve higher localization precision while more fully exploiting the high-quality cleaned representations produced by MSWNet.

\subsection{Other considerations}
\label{sec:others}

% Drift-rate regime and robustness of patch stitching
Our frequency-only tiling is an engineering choice to ensure continuity across patch boundaries, and does not limit the drift-rate range explored in the underlying search. In FAST targeted SETI pipelines, a maximum drift-rate of $\pm4~\mathrm{Hz/s}$ is commonly adopted for hit generation. Typical celestial-mechanical contributions can be much smaller (e.g., at $\nu_0=1.45~\mathrm{GHz}$, Earth-rotation and Earth-orbit terms are $\sim0.15$ and $\sim0.03~\mathrm{Hz/s}$, while a representative exoplanet orbital term can reach $\sim0.57~\mathrm{Hz/s}$; \citealt{2022AJ....164..160T}), so a conservative overlap budget is sufficient.

To avoid erroneous merges in interference-dominated regions, we stitch boundary-truncated segments only when drift direction and morphology are consistent and the overlap region is sparsely populated; otherwise we discard stitching, prioritizing precision over completeness. In manual audits of stitching cases, we found stitching-induced false positives to be more frequent than in the native patch-level outputs, primarily driven by fixed-drift, comb-like RFI: when the line spacing is small, non-contiguous slanted traces can be mistakenly linked across the overlap. Fortunately, detections requiring stitching are rare, and the combination of IoU-based suppression and stitching allows us to tighten the overlap threshold to control such failure modes; in the 803 candidates we inspected, none were created by stitching.

\subsection{Conclusion}
\label{sec:conclusion}

We have presented a wavelet-integrated search pipeline for FAST targeted SETI, aimed at a practical bottleneck in modern technosignature surveys: turning large volumes of ambiguous time--frequency structure into a compact, inspectable, and reproducible candidate stream. In this workflow, MSWNet serves as a structured front end rather than a standalone black-box detector, producing cleaned representations on which lightweight endpoint regression and staged validation can operate.

Applied to the FAST 33-target dataset, the pipeline recovers representative narrowband drifting events from prior FAST analyses and produces a tractable candidate set for expert inspection. The real-data results also highlight an important lesson for ML-assisted SETI: candidate generation and candidate interpretation should remain separate. Narrowband morphology, drift, and single-beam localization are useful filters, but scientific weight still depends on the broader evidence chain, including polarization behavior, cross-target recurrence, ensemble systematics, and multi-beam veto.

Looking forward, the highest-leverage upgrades lie in exploiting raw voltage data from the 19 beams. Cross-correlation across relevant time–frequency windows enables solving for candidate source position, phase consistency, and motion, providing a physical RFI rejection layer that is significantly stronger than current signal-level filtering. The main constraint is timely candidate detection to preserve voltage buffers, but this is fundamentally a software/latency optimization problem. In parallel, endpoint localization can be strengthened via hybridization of the current estimator with physics-driven track fitting, while broader training and validation across diverse RFI regimes would improve cross-survey robustness and backend invariance.

%%%%%%%%%%%%%%%%%%%%%%%%%%%%%%%%%%%%%%%%%%%%%%%%%%%%%%%%%%%%%%%%%%%%%%%%%%%%%%%%
\begin{acknowledgments}
This work was supported by the National Key R\&D Program of China, No.2024YFA1611804 and the China Manned Space Program with grant No. CMS-CSST-2025-A01. This work was supported by the National Natural Science Foundation of China (Nos. 12403004 and 12588202). This work made use of the data from FAST (Five-hundred-meter Aperture Spherical radio Telescope). FAST is a Chinese national mega-science facility, operated by National Astronomical Observatories, Chinese Academy of Sciences. 
\end{acknowledgments}

\facilities{FAST (Five-hundred-meter Aperture Spherical radio Telescope)}

% \appendix

\bibliography{aasjournal}{}
\bibliographystyle{aasjournalv7}

\end{document}